\documentclass[%
11pt,
 amsmath,amssymb,
prb,
]{revtex4-2}

\usepackage{graphicx}% Include figure files
\usepackage{epstopdf}
\usepackage{dcolumn}% Align table columns on decimal point
\usepackage{bm}% bold math
\usepackage{xcolor}
\usepackage{upgreek}
\usepackage{multirow}
\begin{document}

%\preprint{APS/123-QED}

\title{Transport and two-way coupling effect of inertial particles by large-scale and very-large-scale motions in turbulence}
% Force line breaks with \\
%\thanks{A footnote to the article title}%

\author{Guiquan Wang}
% \altaffiliation[Also at ]{Physics Department, XYZ University.}%Lines break automatically or can be forced with \\
\author{David Richter}%
 \email{David.Richter.26@nd.edu}
\affiliation{%
Department of Civil and Environmental Engineering and Earth Sciences, University of Notre Dame, Notre Dame, IN 46556, USA \\
}%

\date{\today}% It is always \today, today,
             %  but any date may be explicitly specified

\begin{abstract}

Direct numerical simulations two-way coupled with inertial particles are used to investigate the particle distribution and two-way coupling effect of low-inertia ($St_{LSM}=0.0625$, $St_{VLSM}=0.009$) and high-inertia ($St_{LSM}=0.475$, $St_{VLSM}=0.069$) particles associated with the large-scale motions (LSMs) and very-large-scale motions (VLSMs) in an open channel flow at a Reynolds number of $Re_\uptau = 550$. One method of filtering the VLSMs from the flow is via artificial domain truncation, which alters the mean particle concentration profile and particle clustering due to the removal of VLSMs from a large domain simulation. In order to exclude possible correlation of the turbulence introduced by a small domain size with periodic boundary conditions, low- and high-pass filtering is performed during the simulation to isolate the particle interaction with different spatial scales. The results show that particle accumulation and turbophoresis are under-predicted without VLSMs, whereas the particle clustering and two-way coupling effects are mainly determined by particle coupling with LSMs. In the inner layer, the elongated streamwise anisotropic particle clustering can be reproduced by particles coupling solely with LSMs for low Stokes number ($St_{LSM}=0.0625$) particles. However, we do not observe similar particle clustering behavior in the outer layer as seen in the full simulation by coupling particles with either LSMs or VLSMs for high Stokes number ($St_{VLSM}=0.069$) particles. This indicates that the organized particle structures are formed by the joint action of LSMs and VLSMs, especially for high Stokes number particles in the outer layer.

\end{abstract}

%\keywords{Suggested keywords}%Use showkeys class option if keyword
                              %display desired
\maketitle

%\tableofcontents
%==============================================================================
\section{Introduction} \label{sec:Introduction}

The physical processes governing small, inertial particles suspended in wall turbulence take place in the transport of the sand dust \cite{dyer1988sand, kok2012physics}, ocean spray \cite{veron2015ocean}, pollutants in the atmospheric boundary layer \cite{guha2008transport}, and the transport of various substances in rivers \cite{nezu2005open}.

In wall turbulence, the large-scale motions (LSMs) play a crucial role in determining the structure and the dynamic process of the entire inner layer \cite{kline1967structure}. The LSMs in the inner layer have characteristic lengths of $\lambda^+_x=O(1000)$ and widths of $\lambda^+_z=O(100)$ in viscous wall units \cite{kline1967structure, jimenez2011cascades}. In the outer layer, the energetic so-called very-large-scale motions (VLSMs) are observed in different wall turbulence flow configurations \cite{Kim1999PoF, del2003spectra, hutchins2007evidence}, and carry $40-65\%$ of the kinetic energy and $30-50\%$ of the Reynolds shear stress \cite{balakumar2007large}. The spanwise wavelength of VLSMs scale as $\lambda_z \geq h$ while their streamwise wavelength is approximately $\lambda_x \geq 10h$ (where $h$ is the boundary layer thickness in turbulent boundary layer or half of the gap size in turbulent channel flow) \cite{Kim1999PoF, del2003spectra, guala2006large, adrian2012coherent}.

These multiscale turbulent structures result in a corresponding wide range of particle-to-fluid time scales which vary as a function of wall-normal height, complicating simple descriptions of particle transport in wall turbulence \cite{fessler1994preferential, pan1995numerical, rouson2001preferential, marchioli2002mechanisms, balachandar2010turbulent, sardina2012wall, richter2013momentum, lee2015modification, wang2019modulation, wang_richter_2019}. In the inner layer, the strongly coherent ejection and sweep motions govern the particle transfer mechanisms \cite{marchioli2002mechanisms}. Conceptually, inertial particles near the wall are swept into low-speed streaks, from whence they are ejected again into the flow \cite{sumer1981particle}, which in turn changes the ejection/sweep intensities \cite{richter2013momentum}, modifying near-wall turbulence monotonically as a function of particle Stokes number \cite{lee2015modification} by modulating the regeneration cycle of LSMs \cite{wang2019modulation}. However, particles can also be trapped underneath the low-speed streaks for a long time \cite{marchioli2002mechanisms} and this causes particle clustering in the near-wall streaks due to turbophoresis \cite{caporaloni1975transfer, reeks1983transport}, which is experimentally observed by \citet{fessler1994preferential}. The characteristic spanwise spacing of particle clustering structures is about $100$ wall units in turbulent Poiseuille flow \cite{bernardini2013effect}.

In the outer layer, it remains a challenge to simulate particle-laden flow in a domain which is long and wide enough to fully resolve the VLSMs, leading to a lack of understanding of inertia particles' transport and clustering with the VLSMs. In order to perform a direct numerical simulation (DNS) accessible Reynolds number and domain size, \citet{bernardini2013effect} use a turbulent Couette flow at $Re_\tau=167$ to study particle clustering in VLSMs inspired by \citet{pirozzoli2011large}, who observed similar inner/outer layer interaction mechanism in low-Reynolds-number turbulent Couette flow as that is in high-Reynolds-number turbulent boundary layers and channels. In their study, \citet{bernardini2013effect} found that the very-large-scale organization of particles with a spacing of $2h$ at one wall linked to the other wall, and organized in the well-known large-scale rows associated with turbulent Couette flow. However, the very-large-scale rows in the turbulent Couette flow are not necessarily the same as VLSMs \cite{avsarkisov2014turbulent}.

Recently, \citet{wang_richter_2019} for the first time to examine the effects of a wide range of particle inertia effect on the VLSMs in an open channel flow at $Re_\tau=550$ and ,$Re_\tau =950$ with a domain size of $L_x=6\pi h$ and $L_z=2\pi h$, which is comparable to the domain size used by \cite{del2003spectra} in single-phase turbulent channel flow. Two distinct particle clustering phenomena appear in the inner layer and outer layer, corresponding to different particle Stokes numbers. One is the well-established particle clustering in near-wall streaks in the inner layer (e.g. $St^+=24.2$) and the other is a new type of organized structure in the outer layer (e.g. $St^+=182$). However, the organized structure in the outer layer is significantly distinct from that previously observed in turbulent Couette flow at $Re_\tau=167$ by \citet{bernardini2013effect}.

\citet{sardina2012wall} studied the artificial domain truncation effect on particle distribution in turbulent channel flow and reported an increase in particle concentration at the wall of up to $20\%$ compared to the small domain at $Re_\uptau=180$. They attributed this difference to the possible correlation of the turbulence and the near-wall particle aggregates in the small domain, which can be excluded in large domain. However, although $Re_\uptau=180$ is not high enough to separate the inner and outer regions, weak VLSMs still exist at this Reynolds number \cite{papavassiliou1997interpretation, toh2005interaction}. As a consequence, it is impossible to exclude the VLSM effect on the particle concentration difference between the large and small domain simulations.

%One of the key challenges in turbulence research is to understand relationships between small and large scales of motion \cite{meneveau2000scale}. DNS coupling with pointwise particles fully resolves the turbulence scale range which is a viable tool for many applications, however, the excessive time-consuming method is still limited to small-scale domain simulation at low to moderate Reynolds numbers . Instead of DNS,  to study ,

In addition to DNS, large eddy simulation (LES) can also be used with Lagrangian tracking to study high Reynolds number, particle-laden flows. Here, the fluid velocity at the particle position is not exactly known, but only a filtered fluid velocity is available \cite{kuerten2016point}. In particle-laden wall turbulent flow, \citet{wang1996large} show that LES predicts the preferential concentration reasonably well both near the wall and along the channel centerline for particles with $St^+=O(10-1000)$ at $Re_\uptau=180,~640$. Later, \citet{marchioli2008some} find that $St^+=O(0.1-100)$ particles underestimate the particle wall accumulation and local segregation for $Re_\uptau=180$. \citet{fede2006numerical} show that particle accumulation is significantly influenced when the particle relaxation time is of the same order or smaller than the subgrid Lagrangian integral time scale measured along particle paths. Therefore, it is necessary to develop an accurate closure model for the inertial particle-subgrid scale interaction in order to predict particle-laden LES \cite{jin2010subgrid}. In this context, most of the particle-laden LES studies have not paid sufficient attention to particles transport by VLSMs, and their relative importance compared to local turbulent structures resolved by DNS. 

%In single-phase flow, by tuning the Smagorinsky constant in a large eddy simulation (LES), \citet{hwang2010self} and \citet{rawat2015self} find a similar self-sustained process exists for VLSMs as the regeneration cycle of LSMs investigated by \citet{hamilton1995regeneration}. They find the scale of VLSMs from LES by quenching the LSMs in the inner layer is similar with full DNS simulation. Therefore, it is necessary to study how inertia particles distribution and clustering behavior by VLSMs and LSMs, which is helpful in further modeling of particle-laden LES in very-large-scale simulations. 

In this work, we first study the truncated domain size effect on the particle distribution, which can primarily isolate LSMs (i.e. exclude VLSMs) from the large domain simulation. However, as discussed above, it is impossible to exclude both the VLSMs effect and the possible artificial domain truncation effects. Therefore as a second method, we isolate LSMs and VLSMs and their one- and two-way coupling effects independently in the same turbulent flow via spatial filtering of the particle advection velocity field at every time step. 

%==============================================================================
\section{Simulation method and parameters} \label{sec:method and parameter}

\subsection{Numerical method} \label{subsec:method}

Direct numerical simulations of the carrier phase are performed for an incompressible Newtonian fluid. A pseudospectral method is employed in the periodic directions (streamwise $x$ and spanwise $z$), and second-order finite differences are used for spatial discretization in wall-normal, $y$ direction. The solution is advanced in time by a third-order Runge-Kutta scheme. Incompressibility is achieved via the solution of a pressure Poisson equation. The fluid velocity and pressure fields are a solution of the continuity and momentum balance equations in Eqs. (\ref{eq:method_conti}) and (\ref{eq:method_momen}), respectively:

\begin{equation}\label{eq:method_conti}
\frac{\partial u_j}{\partial x_j}=0,
\end{equation} 

\begin{equation}\label{eq:method_momen}
\frac{\partial u_i}{\partial t} + u_j\frac{\partial u_i}{\partial x_j}=-\frac{1}{\rho_f} \frac{\partial p}{\partial x_i}+\nu \frac{\partial u_i}{\partial x_j \partial x_j} + \frac{1}{\rho_f} F_i.
\end{equation} 

%%%%%%particles
\noindent Here $u_{i}$ is the fluid velocity, $p$ is the pressure, $F_i$ is the particle feedback force to the carrier phase computed by summing and projecting the particle force to the nearest Eulerian grid points, $\nu$ is the fluid kinematic viscosity, and $\rho_{f}$ is the fluid density. 

Particle trajectories and particle-laden flow dynamics are based on the point-force approximation where the particle-to-fluid density ratio $r \equiv \rho_p/\rho_f \gg 1$ and the particle size is smaller than the smallest viscous dissipation scales of the turbulence. As a consequence of this and the low volume concentrations (a maximum bulk volume fraction of $\overline{\Phi_{V}}$ less than $1 \times 10^{-3}$), only the Schiller-Naumann \citep{schiller1933ber} hydrodynamic drag force is considered. The velocity of particle $n$ is governed by Eq. (\ref{eq:method_drag_f}) and particle trajectories are then obtained from numerical integration of the equation of motion in Eq.  (\ref{eq:method_motion}):

\begin{equation}\label{eq:method_drag_f}
\frac{d u^n_{p,i}}{dt}={f^n_i},
\end{equation} 

\begin{equation}\label{eq:method_motion}
\frac{dx^n_i}{dt}=u^n_{p,i},
\end{equation} 

\noindent where the drag is given by

\begin{equation}\label{eq:method_drag_f_2}
{f^n_i} =\frac{1}{\tau_p}[1+0.15(Re^n _p)^{0.687}] (u^n_{f,i}-u^n_{p,i}).
\end{equation} 

\noindent Here, $\tau _p = \rho _p {d_p} ^2 / 18\mu$ is the Stokes relaxation time of the particle, and the particle Reynolds number $Re^n_p=\mid u^n_{f,i}-u^n_{p,i}\mid d^n_p / \nu $ is based on the magnitude of the particle slip velocity $(u^n_{f,i}-u^n_{p,i})$ and particle diameter $d_{p}^{n}$. In this work, the average $Re^n_p$ is less than $1.0$, which is far smaller than the suggested maximum $Re_{p} \approx 800$ for the Stokes drag correction in Eq. (\ref{eq:method_drag_f}). As a result of the low $Re_{p}$, the correction to the Stokes drag is minimal in this study. Other terms in the particle momentum equation \cite{maxey1983equation} are neglected since they remain small compared with drag when the density ratio $r \gg 1$. In all simulations, particles are initially distributed at random locations throughout the channel. Particle-particle collisions are not taken into consideration, and we exert a purely elastic collision between particles and the lower wall and the free-surface of the open channel flow. Gravity is not included so as to focus specifically on the role of turbulence in particle transport. Validation of the implementation of this code for inertial particles of $St^+=30-2000$ against published numerical and experimental results can be found in \citet{wang2019inertial}.

\subsection{Numerical parameters and domain setup} \label{subsec:parameter}

\begin{table*}
\caption{\label{tab:table1} Parameters of numerical simulations}
\begin{ruledtabular}

\begin{tabular}{clcccccccc}
Type 1 & \multicolumn{1}{c}{ large domain} &       &                   &                       &                    &            &        &            &             \\
\multicolumn{10}{c}{$N_x \times N_y \times N_z = 1024 \times 128 \times 512$}                                                                                         \\
\multicolumn{10}{c}{$L_x \times L_y \times L_z = 6\pi \times 1 \times 2\pi$}                                                                                          \\
\multicolumn{10}{c}{$L^+_x \times L^+_y \times L^+_z = 10367 \times 550 \times 3456$}                                                                                 \\
\multicolumn{10}{c}{$\Delta x^+ \times \Delta y^+ (wall,~surface) \times \Delta z^+=10.1 \times (1,~7.2)  \times 6.75$}                                               \\
Type 2 & \multicolumn{1}{c}{small domain} &         &                   &                       &                    &            &        &            &             \\
\multicolumn{10}{c}{$N_x \times N_y \times N_z = 128 \times 128 \times 128$}                                                                                         \\
\multicolumn{10}{c}{$L_x \times L_y \times L_z = 2.5 \times 1 \times 1.5$}                                                                                          \\
\multicolumn{10}{c}{$L^+_x \times L^+_y \times L^+_z = 1375 \times 550 \times 825$}                                                                                 \\
\multicolumn{10}{c}{$\Delta x^+ \times \Delta y^+ (wall,~surface) \times \Delta z^+=10.7 \times (1,~7.2)  \times 6.45$}                                                            
\\
\multicolumn{10}{c}{}
\\
$Type$ & $Num$                & $\overline{\Phi_m}$ & $\rho_p / \rho_f$ & $\overline{\Phi_v}$   & $N_p$              & $\uptau_p$ & $St^+$ & $St_{LSM}$ & $St_{VLSM}$ \\ 
\\
1      & $case0$              & \multicolumn{8}{c}{Unladen flow}                                                                                                      \\
2      & $case0_{small}$      & \multicolumn{8}{c}{Unladen flow}                                                                                                      \\
\\
1      & $case1$              & $0.14$              & $160$             & $8.75 \times 10^{-4}$ & $7.33 \times 10^6$ & $5.1$      & $24.2$ & $0.0625$   & $0.009$     \\
1      & $case1_{LSM}$        &                     &                   &                       & $7.33 \times 10^6$ &            &        &            &             \\
1      & $case1_{VLSM}$       &                     &                   &                       & $7.33 \times 10^6$ &            &        &            &             \\
2      & $case1_{small}$      &                     &                   &                       & $2.32 \times 10^5$ &            &        &            &             \\
\\
1      & $case2$              & $0.14$              & $1200$            & $1.17 \times 10^{-4}$ & $9.8 \times 10^5$  & $38.2$     & $182$  & $0.475$    & $0.069$     \\
1      & $case2_{LSM}$        &                     &                   &                       & $9.8 \times 10^5$  &            &        &            &             \\
1      & $case2_{VLSM}$       &                     &                   &                       & $9.8 \times 10^5$  &            &        &            &             \\
2      & $case2_{small}$      &                     &                   &                       & $3.1 \times 10^4$  &            &        &            &            
\end{tabular}

\end{ruledtabular}
\end{table*}

The flow configuration of interest is pressure-driven open channel flow. A no-slip condition is imposed on the bottom wall and a shear-free condition is imposed on the upper surface, and such boundary conditions have been proven capable of capturing many of the phenomena (e.g. VLSMs) seen in experiments with shear-free upper boundaries; see \cite{pan1995numerical,Pan1996PoF, adrian2012coherent}. The mesh independence test and single-phase flow validation against \citet{yamamoto2001turbulence} at $Re_\uptau=200$ can be found in \citet{wang_richter_2019}.

An overview of the simulation cases is shown in Table \ref{tab:table1}. The friction Reynolds number is $Re_\tau\equiv u_{\uptau}h/\nu=550$ where $h$ is the depth of the open channel and the particle relaxation time is $\uptau _p \equiv \rho_p d^2/(18\rho_f \nu)$ where $d$ is the particle diameter. The superscript ``+" refers to normalization based on viscous scale, where $\delta_\nu$, $u_{\uptau}$ and $\nu/ u^2_{\uptau}$ correspond to the viscous length scale, velocity scale, and time scale, respectively. 

In the inner layer ($y^+<100$), an autonomous regeneration mechanism maintains the near-wall turbulence (above $y^+=20$), where the characteristic scale of LSMs is roughly $L^{LSM}_y \approx 80$. \citet{wang2017modulation} define a characteristic time scale $\uptau^{LSM}_{f} \sim L^{LSM}_y/max(v'^+ | w'^+)$, which is related to LSMs and is approximately equal to $80$. In the outer layer, the VLSMs nearly extend from the bottom wall to the upper free-surface where the characteristic scale of LSMs can be defined as $L^{VLSM}_y \approx 550$. We similarly define a characteristic time scale $\uptau^{VLSM}_{f} \sim L^{VLSM}_y/max(v' | w')$ related to VLSMs, which approximately equals to $550$. From these, two Stokes numbers are defined for each particle, denoted by $St_{LSM}$ and $St_{VLSM}$. The ratio $d_p/\eta_K$ is maintained at a value of approximately $0.42$, and the particle Reynolds number remains $\mathcal{O}(1)$ or lower. $\overline{\Phi_m}$ is the particle mass concentration and $N_p$ is the total particle number.

In our former work \cite{wang_richter_2019}, low Stokes number particles of $St^+=24.2$ ($St_{LSM}=0.0625$) are more preferentially concentrated in the low-speed regions in the inner layer whereas high Stokes number particles of $St^+=182$ ($St_{VLSM}=0.069$) tend to form distinct clustering structures in the outer layer. Based on these unique dynamics, we choose these two Stokes numbers corresponding to $case1$ and $case2$ in this work to investigate particle transfer by LSMs and VLSMs. 

First, we select two domain sizes for unladen flow ($case0$ and $case0_{small}$), low Stokes number particles ($case1$ and $case1_{small}$) and high Stokes number particles ($case2$ and $case2_{small}$), to test the truncated domain size effect on the particle distribution and clustering effect. The large box has been demonstrated to capture VLSMs in the outer layer and the streamwise turbulent kinetic energy spectrum is nearly unchanged compared with a doubled domain size \cite{wang_richter_2019}. The small box is chosen to capture the LSMs in the inner layer while excluding the VLSMs in the outer layer (this will be shown in Sec \ref{subsec:domainsize}). Then, we artificially couple the particle with LSMs and VLSMs independently in the large domain (corresponding to $case1,2_{LSM}$ and $case1,2_{VLSM}$), in order to investigate particle coupling and transport with LSMs or VLSMs directly (this will be shown in Sec \ref{subsec:couple_filtered_flow}). 
%==============================================================================

\section{Results} \label{sec:Result}

\subsection{Truncated domain size effect} \label{subsec:domainsize}

Particle distribution and transport behavior are determined by the multiscale turbulent structures in wall turbulence, especially in high Reynolds numbers. Therefore, understanding the turbulent structures in single-phase flow is the first priority. As shown by previous studies of domain size effect in single-phase simulations \cite{flores2010hierarchy, hwang2010self, lozano2014effect}, a proper minimum domain size is important to get `healthy' turbulence (obtaining accurate one-point statistics) in numerical simulations, in order to avoid the constraints of the flow structures due to the effect of the periodic boundary condition. In the viscous and buffer layers, \citet{jimenez1991minimal} use a minimal box with size $L^+_x=300-600$ and $L^+_z=80-160$ in order to isolate the wall-attached structures. The low-order turbulence statistics are in good agreement with experiments in the near-wall region, which is due to the fact that VLSMs carry little Reynolds stress near the wall in full simulations and are largely independent from the autonomous LSMs \cite{jimenez2004large}. Furthermore, \citet{hamilton1995regeneration} use this concept to study the dynamics regeneration cycle of LSMs in the inner layer.

In the logarithmic and outer regions, \citet{flores2010hierarchy} show that similar minimal boxes exist for the logarithmic and outer layers of turbulent channels, but the size ($L_x=6h$ and $L_z=3h$) is much larger than in \citet{jimenez1991minimal}. Recently, \citet{lozano2014effect} demonstrated that the domain size of $L_x=2 \pi h$ and $L_z= \pi h$ is large enough to reproduce the one-point statistics of larger boxes at $Re_\uptau=547-4050$. Meanwhile \citet{hwang2010self} shows that the self-sustaining nature of VLSMs is maintained only if the streamwise and spanwise box sizes are larger than the minimal values $L_x=3h$ and $L_z=1.5h$ at $Re_\uptau=550$.

In particle-laden flow, \citet{sardina2012wall} use a smaller domain size of $L_x=4 \pi h$ ($L^+_x=2260$) and $L_z=4 \pi/3 h$ ($L^+_x=754$) and compare to a larger domain size of $L_x=12 \pi h$ and $L_z=4 \pi h$ for simulating particle-laden flow at a low $Re_\uptau=180$. The small domain size is larger than \citet{hwang2010self} and \citet{lozano2014effect} in outer units, which indicates that this domain size is not short enough to exclude the VLSMs in the outer layer. On the contrary, it is still not long enough to capture decorrelated LSM signatures in the streamwise direction in the inner layer. Therefore with this configuration, it is hard to separate the effect of possible correlation of the turbulence in the inner layer and the influence of VLSMs in both the inner and outer layers. 

\subsubsection{Energy spectrum} \label{subsubsec:energy_spectrum}

%====================Figure================================================
\begin{figure*}[hbt!]
\includegraphics[width=15cm,trim={0cm 0cm 0cm 0cm}, clip]{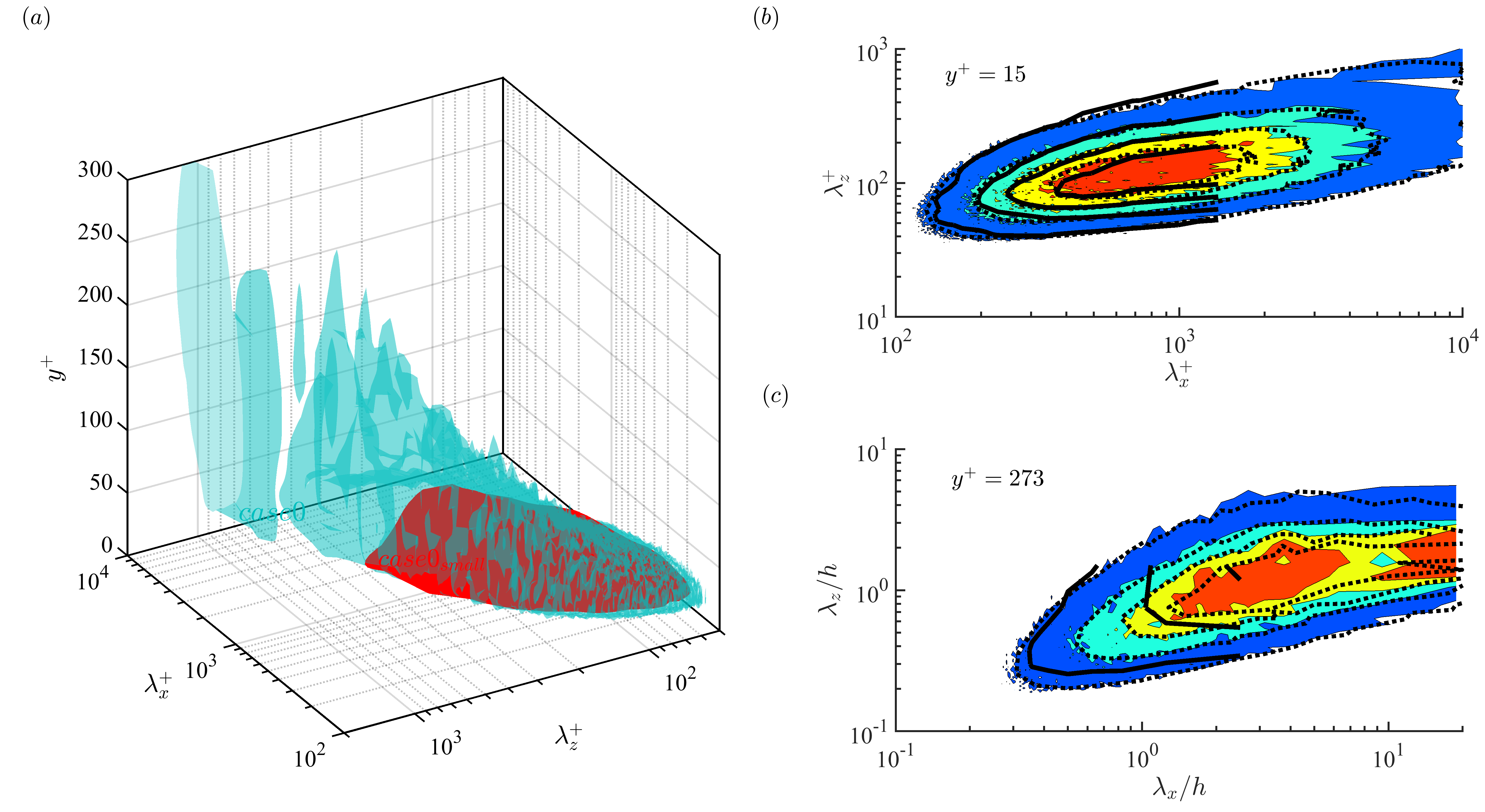} 
\caption{\label{fig:Euu} Premultiplied two-dimensional energy spectrum $k_xk_z\Phi_{u'u'}/u_\uptau ^2$ as a function of $\lambda_x$ and $\lambda_z$ for $case0$ and $case0_{small}$, $(a)$ in the wall-normal direction $y$, isosurface of $0.1$ times the maximum value of the unladen flow is illustrated. $(b)$ and $(c)$ refer to $y^+=15$ and $y^+=273$, respectively. In $(b)$ and $(c)$, the filled contours represent the large domain, lines are from the small domain, and dotted lines are from \citet{del2003spectra}.}
\end{figure*}
%==========================================================================

For the truncated simulations, we choose a domain size of $L_x, L_z=2.5h, 1.5h$ to exclude the VLSMs in the outer layer, corresponding to $L^+_x, L^+_z=1375, 825$ in wall viscous units. The streamwise extent cannot exclude the correlation of LSMs in the inner layer. The premultiplied, two-dimensional energy spectrum of streamwise velocity, $k_xk_z\Phi_{u'u'}$ where $\Phi_{u'u'}=\langle{\hat{u'}(k_x,k_z,y)\hat{u'}^*(k_x,k_z,y)} \rangle$, is shown in Fig. \ref{fig:Euu} for $Re_{\uptau}=550$ ($\hat{u'}$ is the Fourier coefficient of $u'$, $k_x$ is the streamwise wavenumber, and $k_z$ is the spanwise wavenumber). In Fig. \ref{fig:Euu}(a), we can qualitatively see that the small domain ($case0_{small}$) generally well-captures the turbulent structures in the inner layer ($y^+<100$). However, the VLSMs in the inner and outer layers are completely lacking in the small domain simulation (by design). The cross-section at $y^+=15$ and $y^+=273$ compared between $case0$, $case0_{small}$, and results from \citet{del2003spectra} are shown in Figs. \ref{fig:Euu}(b) and (c), respectively. The contour of $k_xk_z\Phi_{u'u'}$ for the large domain agrees well with the results from \citet{del2003spectra}. In the inner layer, the energetic LSMs in $case0_{small}$ are nearly the same as in $case0$, whereas the tail of the spectrum (i.e. $\lambda_x>5h$ in Fig. \ref{fig:Euu}(c)) represents for deep $u-$modes \citep{del2003spectra} or VLSM footprints \citep{hutchins2007evidence}. In the outer layer, the premultiplied two-dimensional energy spectrum of $case0_{small}$ is significantly different from $case0$, which indicates VLSMs are not captured in the smaller domain simulation, as expected.

%====================Figure================================================
\begin{figure*}
\includegraphics[width=15cm,trim={0cm 0cm 0cm 0cm}, clip]{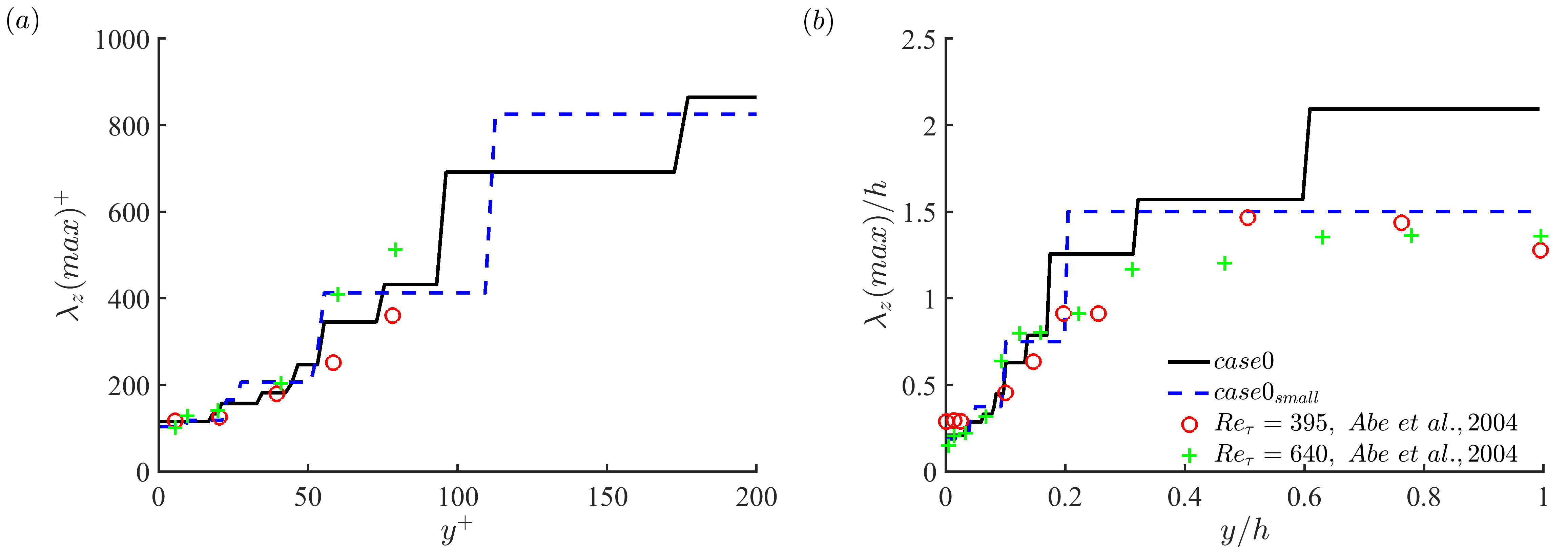}
\caption{\label{fig:Lambda_z_max}Spanwise wavelengths of the most energetic structures obtained from the premultiplied energy spectra of $u'$ for $case0$ and $case0_{small}$ at $Re_\uptau=550$ compared with results of \citet{abe2004very} at $Re_\uptau=395$ and $640$ in channel flow. $(a)$ in wall unit; $(b)$ in outer unit.}
\end{figure*}
%==========================================================================

The spanwise wavelengths $\lambda_{z,max}$ of the most energetic structures obtained from the 1-D premultiplied energy spectra of $u'$ for $case0$ and $case0_{small}$ are shown in Fig. \ref{fig:Lambda_z_max}. Results of \citet{abe2004very} at $Re_\uptau=395$ and $640$ in turbulent channel flow are plotted as well for comparison. In general, the scale of $\lambda_{z,max}$ increases with the wall-normal height and the scale of $\lambda_{z,max}$ in open channel flow is wider than it is in channel flow \cite{abe2004very}. In the inner layer as shown in Fig. \ref{fig:Lambda_z_max}(a), the scale of $\lambda_{z,max}$ is nearly the same between $case0$ and $case0_{small}$ and agrees well with the results of \citet{abe2004very}, which tends to be longer at a higher Reynolds number compared to a lower Reynolds number. In the outer layer above $y=0.6$ as shown in Fig. \ref{fig:Lambda_z_max}(b), the development of $\lambda_{z,max}$ in $case0_{small}$ is constrained by the limited spanwise domain size. The scale of $\lambda_{z,max}$ is wider in $case0$ than $case0_{small}$ and results from \citet{abe2004very}. 

\subsubsection{Reynolds stress} \label{subsubsec:Reynolds_stress}

%====================Figure================================================
\begin{figure*}
\includegraphics[width=14.5cm,trim={0cm 0cm 0cm 0cm}, clip]{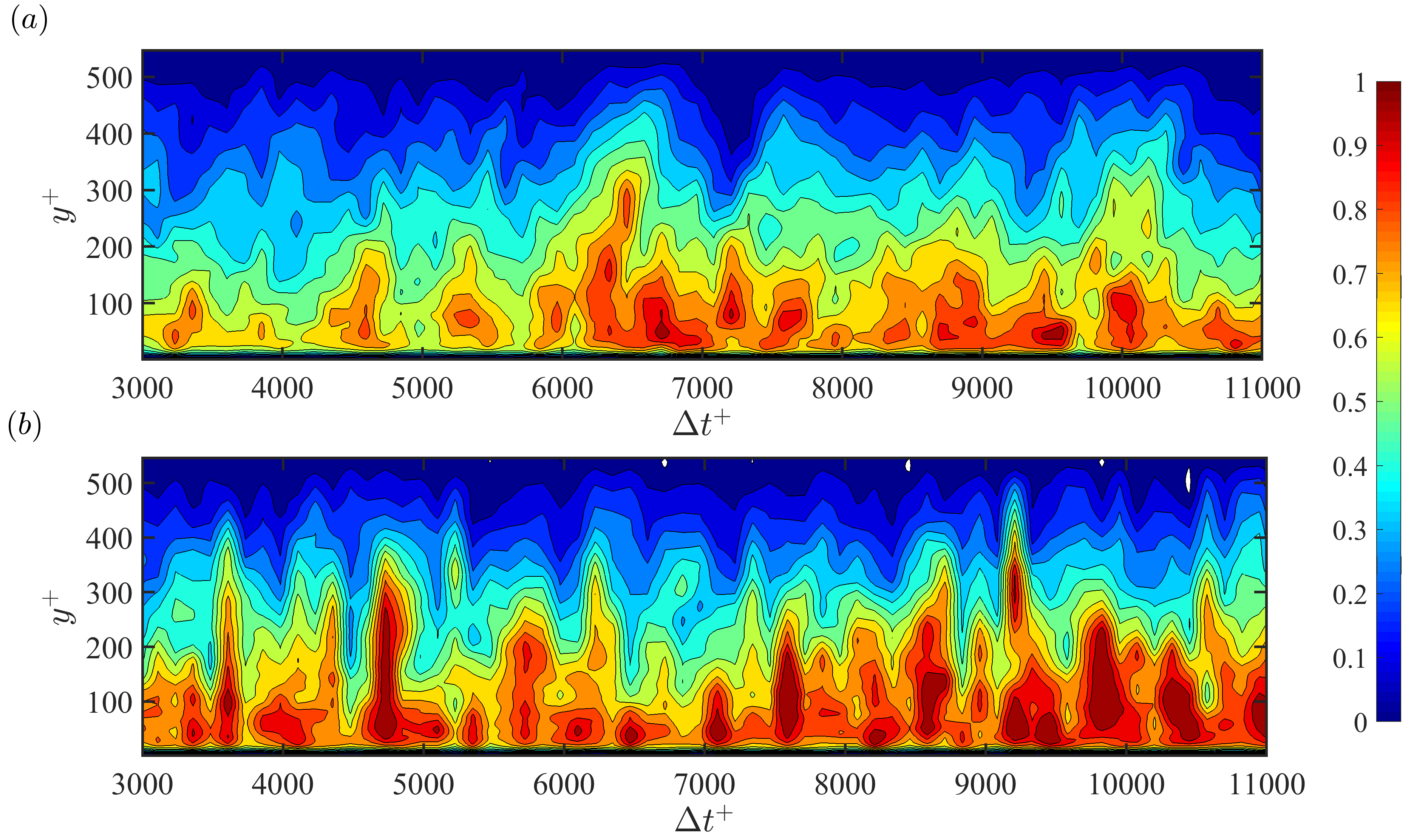}
\caption{\label{fig:shear_stress_time}Evolution in time of shear Reynolds stress $-\overline {u'v'}(y,t)$ averaged over the homogeneous directions and scaled by wall units ($u_\uptau ^2$). (a) Small domain simulation, $case0_{small}$; (b) large domain simulation ($case0$), averaged over same area as small domain.}
\end{figure*}
%==========================================================================

\citet{kline1967structure} find that the dominance of the streaks is not only confined to the inner layer, but also directly or indirectly affect the outer layer. The ejections appear to account for most of the Reynolds shear stress ($-\overline {u'v'}$) \cite{kline1967structure} and the observed intermittent bursting periods contribute almost all of turbulent kinetic energy production ($-\overline {u'v'}dU/dy$) \cite{kim1971production}. By using quadrant analysis, \citet{wallace2016quadrant} shows that the ejection and sweep quadrants make the largest contribution to the Reynolds stress. The intermittent turbulent structures represent the regeneration cycle process of LSMs in the inner layer and has been suggested as a formation mechanism for the organized VLSMs in the outer layer \cite{Kim1999PoF}. For the present simulations, the temporal evolution of horizontally-averaged Reynolds shear stress is shown in Fig. \ref{fig:shear_stress_time}. Compared with the two other configurations (i.e. channel flow and Couette flow) \cite{wang_abbas_climent_2018}, the strongest shear stress bursts in the open channel flow are located close to the bottom wall whereas they are weak close to the free surface. The shear stress bursts are stronger in the larger domain as compared to the small domain across the wall-normal height. As shown in Fig. \ref{fig:Euu}(b) and Fig. \ref{fig:Lambda_z_max}(a), the energetic structures in the small domain simulation are nearly the same as they are in the large domain simulation within the inner layer. However, comparing Fig. \ref{fig:shear_stress_time}(a) with \ref{fig:shear_stress_time}(b), the Reynolds shear stress is higher in large domain than it is in small domain, due to the presence of VLSMs (the long tail in Fig. \ref{fig:Euu}(b) and (c)). In the outer layer, the VLSMs cannot be captured in the small domain simulation leading to a weaker Reynolds shear stress in the small domain simulation (Fig. \ref{fig:shear_stress_time}(a)).

%====================Figure================================================
\begin{figure*}
\includegraphics[width=16cm,trim={0cm 0cm 0cm 0cm}, clip]{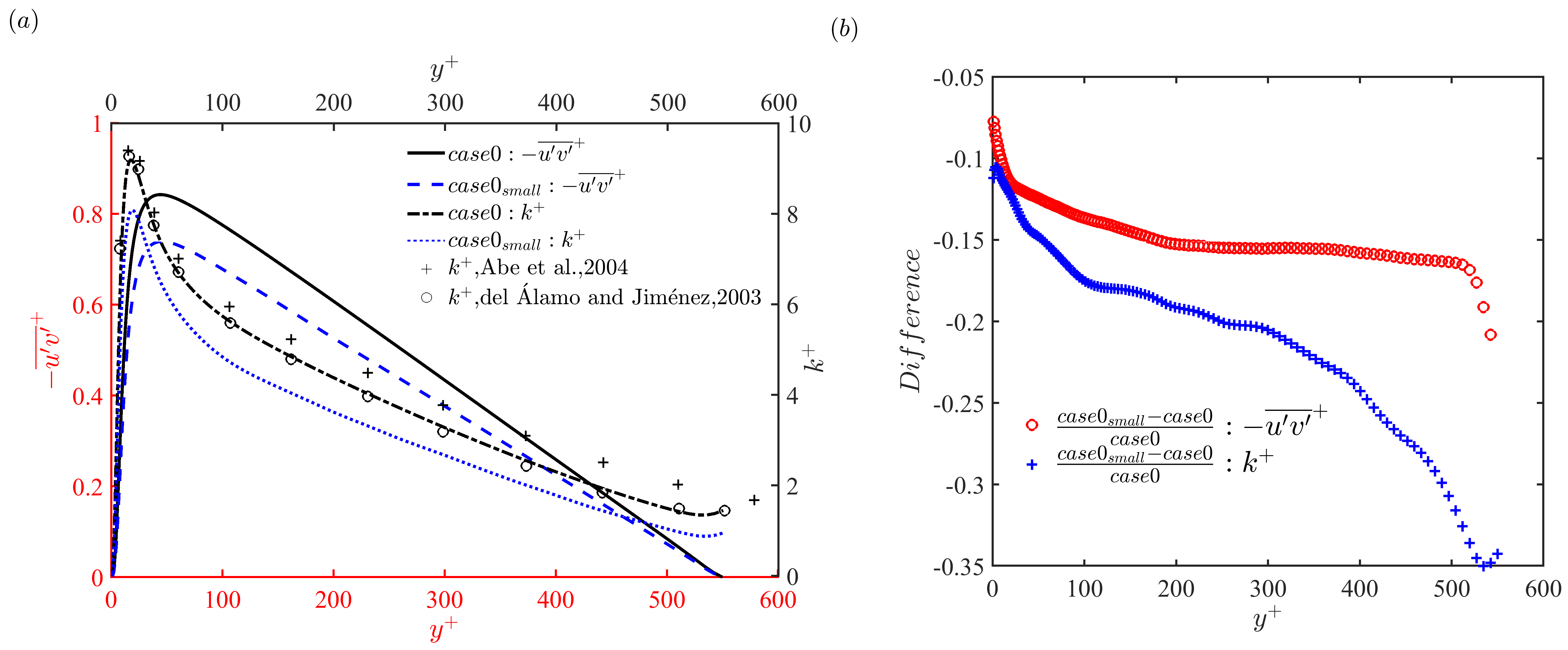}
\caption{\label{fig:uv_TKE} $(a)$ The Reynolds shear stress ($-\overline {u'v'}$) and turbulent kinetic energy ($k$) as a function of wall-normal height, normalized by wall unit ($u_\uptau ^2$). Published results ($k$) of \citet{abe2004very} for $Re_\uptau = 640$ and \citet{del2003spectra} for $Re_\uptau = 550$ in turbulent channel flow are plotted. $(b)$ The difference of $-\overline {u'v'}$ and $k$ between small domain simulation with large domain simulation, normalized by the results of large domain simulation.}
\end{figure*}
%==========================================================================

Furthermore, the temporal average of the Reynolds shear stress in Fig. \ref{fig:shear_stress_time} is shown in Fig. \ref{fig:uv_TKE}(a), accompanied by the normalized turbulent kinetic energy ($k^{+}$). As a comparison, published results of $k$ by \citet{abe2004very} for $Re_\uptau = 640$ and \citet{del2003spectra} for $Re_\uptau = 550$ in a large domain, turbulent channel flow are plotted as well. We see that the turbulent kinetic energy is nearly the same between large simulation $case0$ with the results of \citet{del2003spectra} at same Reynolds number, which is lower than the results of \citet{abe2004very} for higher Reynolds number. Comparing $case0$ with $case0_{small}$, both the Reynolds shear stress and turbulent kinetic energy are higher in the large domain simulation than in the small domain simulation. The difference is shown in Fig. \ref{fig:uv_TKE}(b), which monotonically increases in the wall-normal direction. The Reynolds shear stress (turbulent kinetic energy) difference is around $10\%$ ($15\%$) in the inner layer whereas increases to $15\%$ ($20-35\%$) in the outer layer. The trend is similar but quantitatively smaller than previously observed by \citet{balakumar2007large}, due to the lower Reynolds number. 

\subsubsection{Particle distribution} \label{subsubsec:particle_distribution_small_large}

%====================Figure================================================
\begin{figure*}
\includegraphics[width=15.5cm,trim={0cm 0cm 0cm 0cm}, clip]{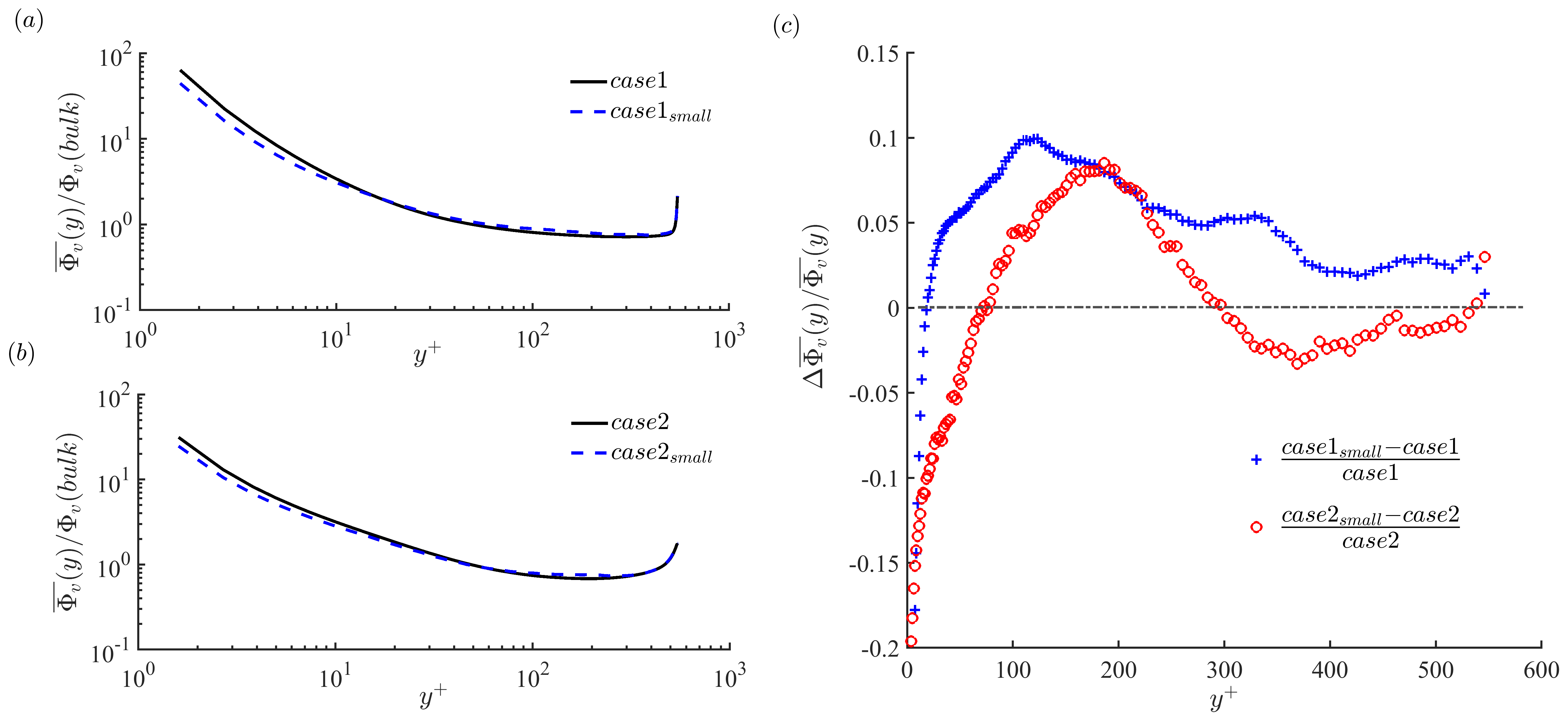}
\caption{\label{fig:Concentration_small_large} Mean particle volume concentration in wall-normal direction, scaled by the bulk value. $(a)$ Low Stokes number, comparison between $case1_{small}$ and $case1$; $(b)$ High Stokes number, comparison between $case2_{small}$ and $case2$. $(c)$ The difference between small domain simulation with large domain simulation, normalized by the results of large domain simulation.}
\end{figure*}
%==========================================================================

We now turn our attention to the particle distributions in the large and small domains, with emphasis on the effect of truncating the VLSM signatures in the small domain. Mean particle volume concentrations for the small domain and large domain simulations are shown in Fig. \ref{fig:Concentration_small_large}. For both low and high Stokes number particles in Figs. \ref{fig:Concentration_small_large}(a) and (b) respectively, there are fewer particles in the near-wall region of the small domain simulation than there are in the large domain simulation, while the opposite trend is observed in the outer region. The difference of mean particle volume concentration is shown in Fig. \ref{fig:Concentration_small_large}(c). An increase in particle concentration near the wall is found in the large domain simulation, up to $20\%$ larger than the small domain. This indicates that the turbophoretic effect is enhanced in the large domain simulation. On the other hand, the region of the lower particle concentration in the small domain expands from the near-wall region to the outer region with increase of the Stokes number, which is likely due to the high-inertia particles ($St_{VLSM}=0.069$) more preferably responding to the VLSMs compared to the low inertial particles ($St_{VLSM}=0.009$). The observed differences due to the truncated domain size effect are similar as previously observed by \citet{sardina2012wall}.

%====================Figure================================================
\begin{figure}
\includegraphics[width=10cm,trim={0cm 0cm 0cm 0cm}, clip]{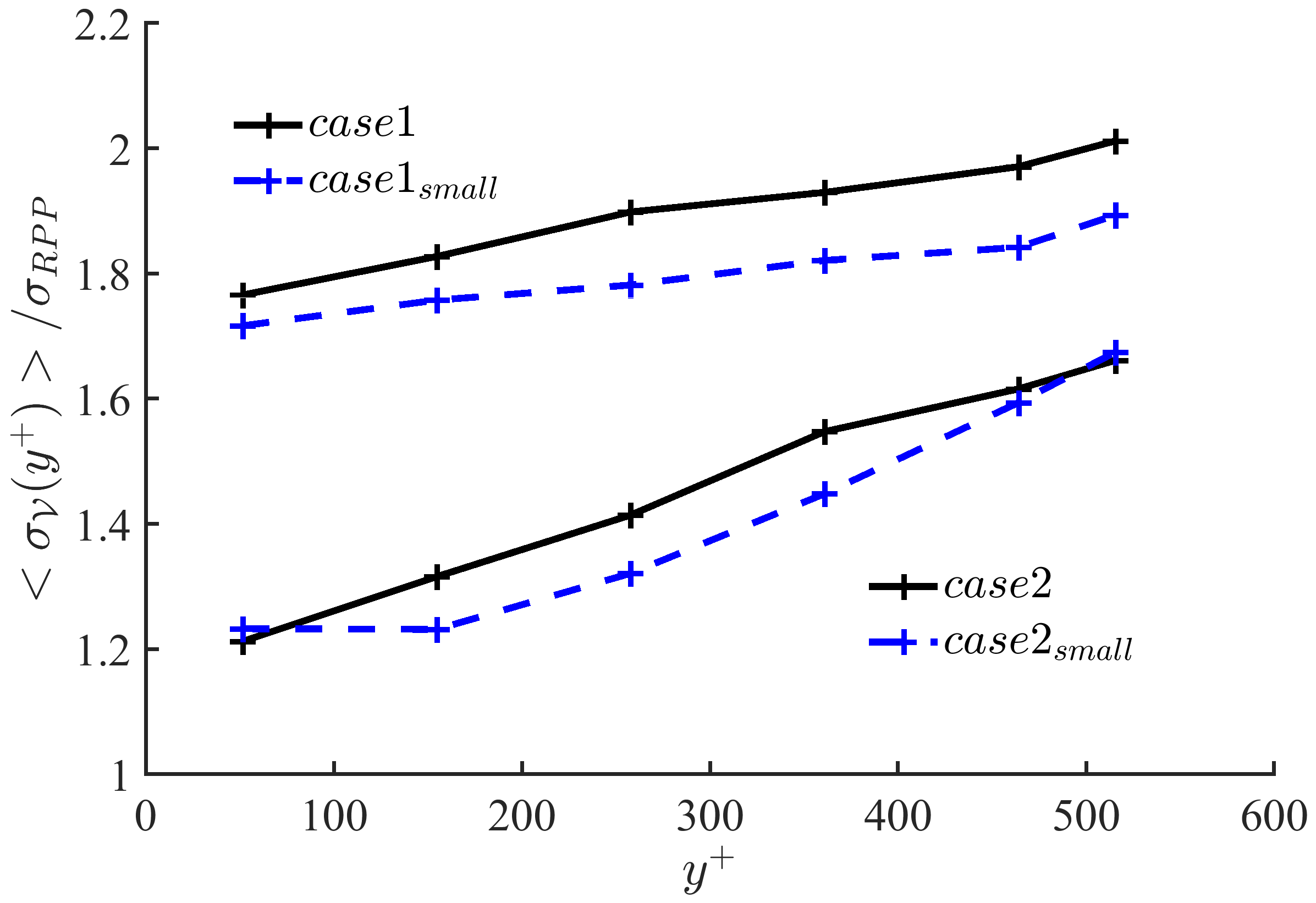}
\caption{\label{fig:voronoi_small_large}Standard deviation of the normalized Vorono\"i area $\sigma_{\mathcal{V}}$, normalized by that of a random Poisson process, $\sigma_{RPP}$, as a function of height in wall-normal direction of low and high Stokes numbers in small domain and large domain.}
\end{figure}
%==========================================================================

As proposed by \citet{monchaux2010preferential,monchaux2012analyzing}, the Vorono\"i diagram can be efficiently used to identify and quantify particle clusters. The standard deviation of the distribution of Vorono\"i areas is directly linked to the level of clustering. For the present study, the instantaneous particle locations are analyzed in six slabs with thicknesses of $2d_p$ at multiple wall-normal distances. Fig. \ref{fig:voronoi_small_large} displays the standard deviation ($\sigma_\mathcal{V}$) of the distribution of the normalized Vorono\"i area $\mathcal{V}=A/\overline{A}$, where the inverse of the average Vorono\"i area $\overline{A}$ indicates the mean particle concentration. $\sigma_\mathcal{V}$ is scaled by the standard deviation of a random Poisson process (RPP; $\sigma_{RPP}=0.52$), which would be expected if particles were randomly distributed. The ratio $\sigma_\mathcal{V}/\sigma_{RPP}$ exceeding unity indicates that particles are accumulating in clusters as compared to truly randomly distributed particles. In the inner layer ($y^+=50$), the particle accumulation is slightly different between the small domain with the large domain. Away from the wall in the outer layer ($150 \leq y^+ \leq 457$), particle preferential accumulation is higher in the large domain simulation than it is in the small domain simulation. Near the free surface, the particle clustering is nearly same for high Stokes number in two different domain simulations whereas it is still higher in large domain than in small domain for low Stokes number. 

Based on the analysis of particle distribution and preferential accumulation, in the inner layer, the particle concentration increases up to $20\%$ due to the influence of VLSMs and the effect of correlated LSMs. However, in the outer layer, particles tend to form a strong clustering due to the influence of VLSMs.

\subsection{Particles coupled with a filtered flow field} \label{subsec:couple_filtered_flow}

As we have discussed in the introduction (Sec. \ref{sec:Introduction}), for low to moderate Reynolds numbers, it is contradictory to have a domain size large enough to decorrelate the LSMs in the streamwise direction within the inner layer, but small enough to exclude the formation and maintenance of VLSMs in the outer layer. As a consequence, both effects together result in the truncated domain size effects discussed in Sec. \ref{subsec:domainsize}. Here, we instead retain the large domain size, but via spatial filtering, only allow particles to interact with specific scales of the turbulent flow.

\subsubsection{Filtered fluid velocity} \label{subsubsec:Filtered_LSM_VLSM}

%====================Figure================================================
\begin{figure*}
\includegraphics[width=16cm,trim={0cm 0cm 0cm 0cm}, clip]{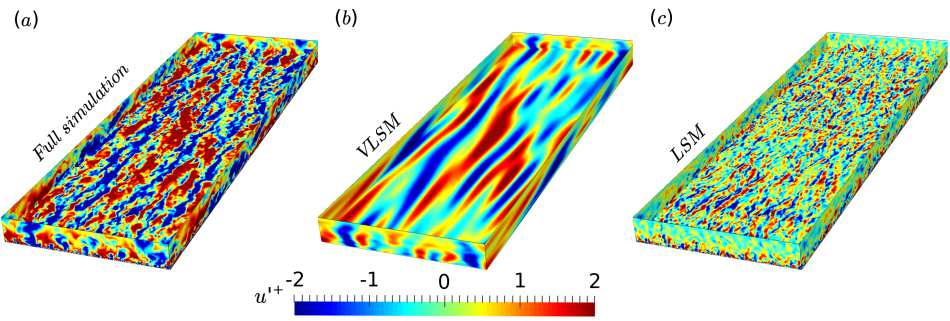}
\caption{\label{fig:visu_LSM_VLSM} Instantaneous contours of streamwise velocity fluctuation on a wall-parallel plane at $y^+=100$ (and domain boundary walls) in single-phase flow ($case0$), normalized by $u_\tau$. $(a)$ Full simulation containing all modes; $(b)$ The same flow field but only associated with VLSMs, containing modes with $\lambda_x > 5h,~\lambda_z > 0.75h$; $(c)$ The same flow field but only associated with LSMs, containing modes with $\lambda_x < 5h,~\lambda_z < 0.75h$.}
\end{figure*}

In this section, we use an artificial coupling technique between selected scales of turbulent structures (i.e. LSMs versus VLSMs) with inertial particles to isolate the LSMs' and VLSMs' role in particle transport behavior and two-way coupling. The filtered fluid velocity field for LSMs and VLSMs (only used to couple with particles), $\widetilde{\mathbf{u}}$ is computed as 

\begin{equation}\label{eq:filtered_flow}
\widetilde{\mathbf{u}}(x,y,z,t) = \mathcal{F}^{-1}
\left\{\begin{matrix}
 \hat{\mathbf{u}}(\lambda_x,y,\lambda_z,t), & if ~ [\lambda_x, \lambda_z] \in ~LSMs~or~VLSMs\\ 
  0, & otherwise\\
\end{matrix}\right.
\end{equation}

\noindent where $\mathcal{F}^{-1}$ is the inverse Fourier transform, $\hat{\mathbf{u}}(\lambda_x,y,\lambda_z,t)$ is the $2D$ Fourier transform of the fluid velocity $\mathbf{u}(x,y,z,t)$ in the two homogeneous directions at every plane in the wall-normal direction at every time step, and the $\lambda_x$ and $\lambda_z$ are the streamwise and spanwise wavelengths, respectively. We define the length scale of the LSMs as $\lambda_x < 5h,~\lambda_z < 0.75h$ and the VLSMs as $\lambda_x > 5h,~\lambda_z > 0.75h$ in Fourier space, respectively, in accordance to that used by \citet{del2003spectra}.

The instantaneous streamwise velocity fluctuation ($u'$) field on a wall-parallel plane at $y^+=100$ and sidewalls is shown in Fig. \ref{fig:visu_LSM_VLSM}(a), and the corresponding spectral information and turbulent kinetic energy can be seen in \citet{wang_richter_2019}. Obviously, the multiscale and turbulent field is composed of both large-scale and very-large-scale motions. By applying Eq. \ref{eq:filtered_flow}, the instantaneous velocity fields of the LSMs and VLSMs can be isolated from full simulation at the same time step as shown in Figs. \ref{fig:visu_LSM_VLSM}(b) and (c), respectively. The streamwise elongated VLSMs are characterized by alternating low-speed and high-speed regions in the spanwise direction, extending from the bottom wall to the free surface in the wall-normal direction (Fig. \ref{fig:visu_LSM_VLSM}(b)). The isolated LSMs are also elongated in the steamwise direction with alternating low-speed and high-speed regions (Fig. \ref{fig:visu_LSM_VLSM}(c)), similar as the VLSMs. Furthermore, there are multiple and decorrelated LSMs contained within the large domain simulation. However, the domain size seems to be not long enough to fully decorrelate the VLSMs in the streamwise direction. 

At every Runge-Kutta substep, inertial particles are conditionally coupled with the filtered flow field $\widetilde{\mathbf{u}}$ based on Eq. \ref{eq:filtered_flow}, representing either LSMs (simulations $case1,2_{LSMs}$) or VLSMs (simulations $case1,2_{VLSMs}$). 

%%====================Figure================================================
%\begin{figure*}
%\includegraphics[width=14cm,trim={0cm 0cm 0cm 0cm}, clip]{./Reynolds_shear_stress_2.png}
%\caption{\label{fig:wide}Use the figure* environment to get a wide
%figure that spans the page in \texttt{twocolumn} formatting.}
%\end{figure*}

\subsubsection{Particle distribution} \label{subsubsec:distribution_LSM_VLSM}

%====================Figure================================================
\begin{figure*}
\includegraphics[width=16cm,trim={0cm 0cm 0cm 0cm}, clip]{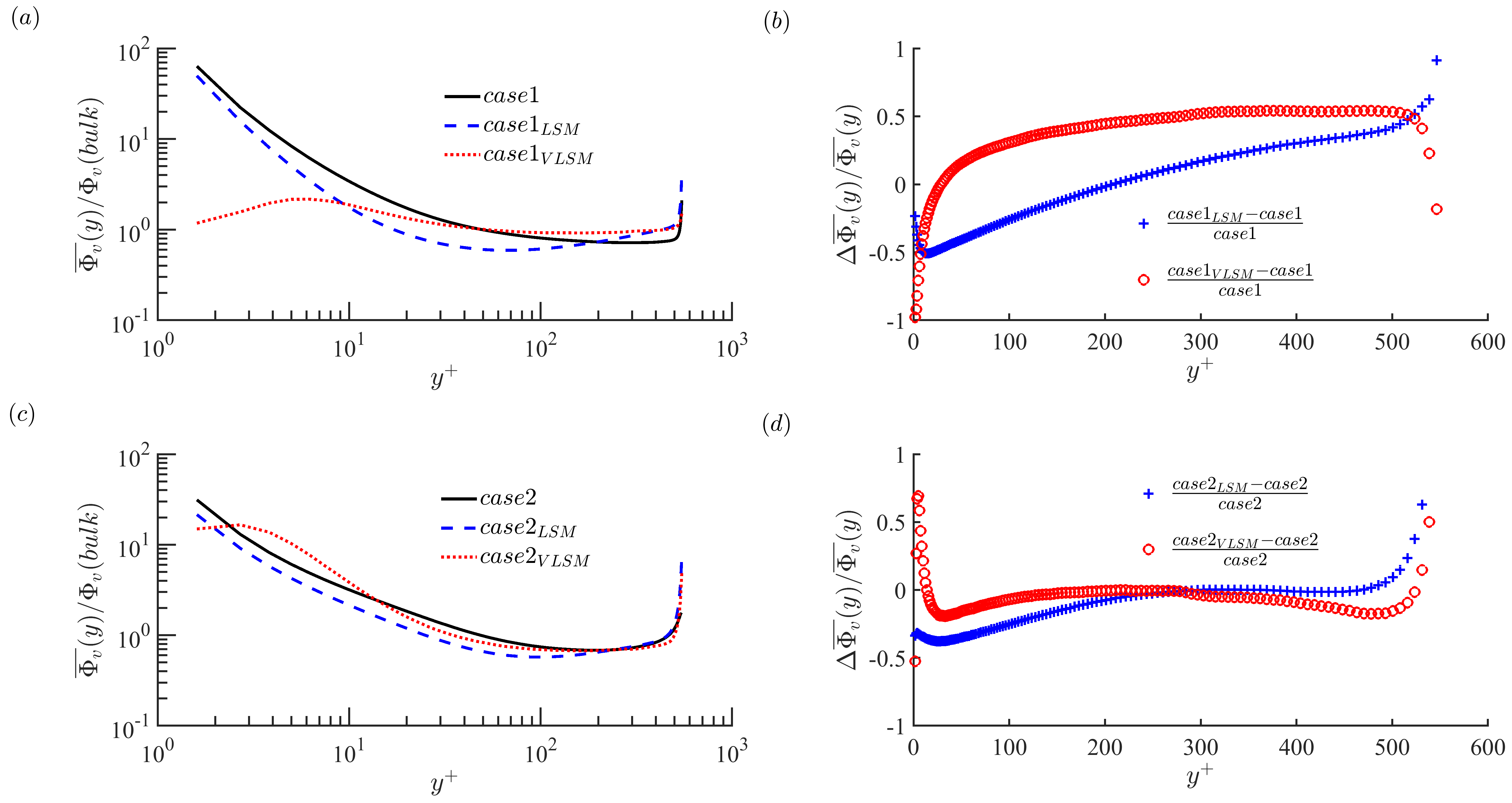}
\caption{\label{fig:Concentration_LSM_VLSM} Mean particle volume concentration in wall-normal direction, compared between full simulation with artificial coupling LSMs or VLSMs, scaled by the bulk value. (a, b) Low Stokes number, comparison between $case1$, $case1_{LSM}$ and $case1_{VLSM}$; (c, d) High Stokes number, comparison between $case2$, $case2_{LSM}$ and $case2_{VLSM}$. (b, d) The difference between particle-laden in LSMs or VLSMs with full simulation, normalized by the results of full simulation.}
\end{figure*}

The mean particle volume concentrations of particles coupled with the filtered velocity fields are compared to the full simulation in Fig. \ref{fig:Concentration_LSM_VLSM}; for low Stokes number in Figs. \ref{fig:Concentration_LSM_VLSM}(a, b) and for high Stokes number in Figs. \ref{fig:Concentration_LSM_VLSM}(c, d). For both low and high inertial particles coupled with the LSMs ($case1,2_{LSM}$), the wall-normal particle concentration profile has a similar shape compared to the full simulations ($case1,2$). Quantitatively, compared with the full simulations, $case1,2_{LSM}$ under-predict (less than $\sim 50\%$) the particle concentration in the region of $y^+ \leq 200$ whereas $case1,2_{LSM}$ over-predict (less than $\sim 50\%$) the particle concentration in the region of $y^+ \geq 200$. The trend is similar to that observed in the small domain size simulation discussed in Sec. \ref{subsubsec:particle_distribution_small_large}. From this we confirm that the particle concentration and the effects turbophoresis are truly under-predicted when VLSMs are absent in numerical studies.

At the same time, however, particle transport behavior by VLSMs is distinctly different for both the low and high inertia particles compared with the full simulations. For low-inertia particles as shown in Figs. \ref{fig:Concentration_LSM_VLSM}(a, b), the particle wall-normal concentration profile of $case1_{VLSM}$ is flatter than it is in full simulation due to the low particle response time scale compared to the time scale of the VLSMs. Quantitatively, $case1_{VLSM}$ seriously under-predicts (even more than $100\%$ in magnitude) the particle concentration in the near-wall region of $y^+ \leq 40$ whereas it over-predicts (less than $50\%$ in magnitude) the particle concentration in the region of $y^+ \geq 40$. For high-inertia particles as shown in Figs. \ref{fig:Concentration_LSM_VLSM}(c, d), the particle concentration profile of $case2_{VLSM}$ is generally similar as it is in full simulation $case2$ except very close to the wall $y^+ \leq 3$. Quantitative comparison indicates that $case2_{VLSM}$ over-predicts (even more than $50\%$  in magnitude) the particle concentration in the near-wall and near surface regions ($3\leq y^+ \leq 13$ and $y^+ \geq 530$) whereas $case2_{VLSM}$ under-predicts the particle concentration in the very-near-wall and very-near-surface regions ($y^+ \leq 3$). At most of the wall-normal region ($13 \leq y^+ \leq 530$), the particle concentration profile agrees fairly well with the full simulation. This all suggests that high-inertia particles are largely transported by VLSMs, while low-inertia particle distribution is dominated by LSMs in the inner layer whereas it is determined by both LSMs and VLSMs in the outer layer.

%====================Figure================================================
\begin{figure*}
\includegraphics[width=16cm,trim={0cm 0cm 0cm 0cm}, clip]{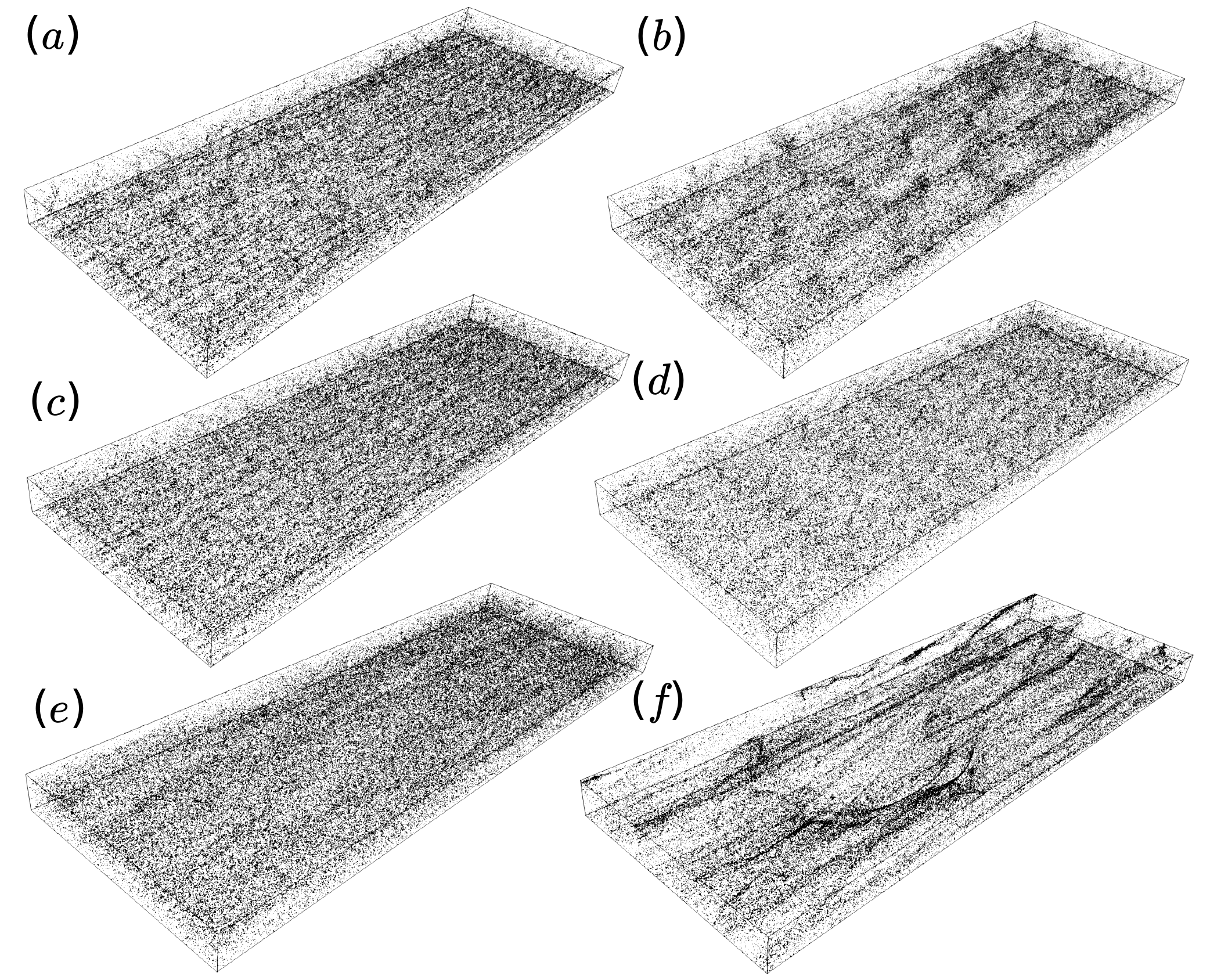}
\caption{\label{fig:Location_LSM_VLSM} Instantaneous snapshots of particle locations (black dots). (a, c, e) Low Stokes number particles; (b,d,f) high Stokes number particles. (a) $case1$; (b) $case2$; (c) $case1_{LSM}$; (d) $case2_{LSM}$; (e) $case1_{VLSM}$; (f)  $case2_{VLSM}$.}
\end{figure*}

Fig. \ref{fig:Location_LSM_VLSM} presents instantaneous snapshots of particle locations (black dots) for the same cases. Compared to Fig. \ref{fig:Location_LSM_VLSM}(a) for the full simulation ($case1$), here we see particles tending to distribute randomly when they only couple with VLSMs ($case1_{VLSM}$) for low Stokes number as shown in Fig. \ref{fig:Location_LSM_VLSM}(e), while LSM coupling with particles can capture the particle accumulation close to the wall as shown in Fig. \ref{fig:Location_LSM_VLSM}(c) (these trends will be confirmed later). The particle response to LSMs and VLSMs is significantly different for high-inertia particles compared with low-inertia particles. Fig. \ref{fig:Location_LSM_VLSM}(b) shows that for the full simulation laden with high inertial particles ($case2$), there are two different clustering structures: the streamwise elongated particle streaks in the inner layer and $3D$ spatial clustering of particles in the outer layer. Particles coupled with the LSMs ($case2_{LSM}$ in Fig. \ref{fig:Location_LSM_VLSM}(d)) form alternating particle clusters in the outer layer, while particles coupled with the VLSMs ($case2_{VLSM}$ in Fig. \ref{fig:Location_LSM_VLSM}(f)) form  elongated, anisotropic structures in the outer layer which are observed in full simulation.

%====================Figure================================================
\begin{figure*}
\includegraphics[width=16cm,trim={0cm 0cm 0cm 0cm}, clip]{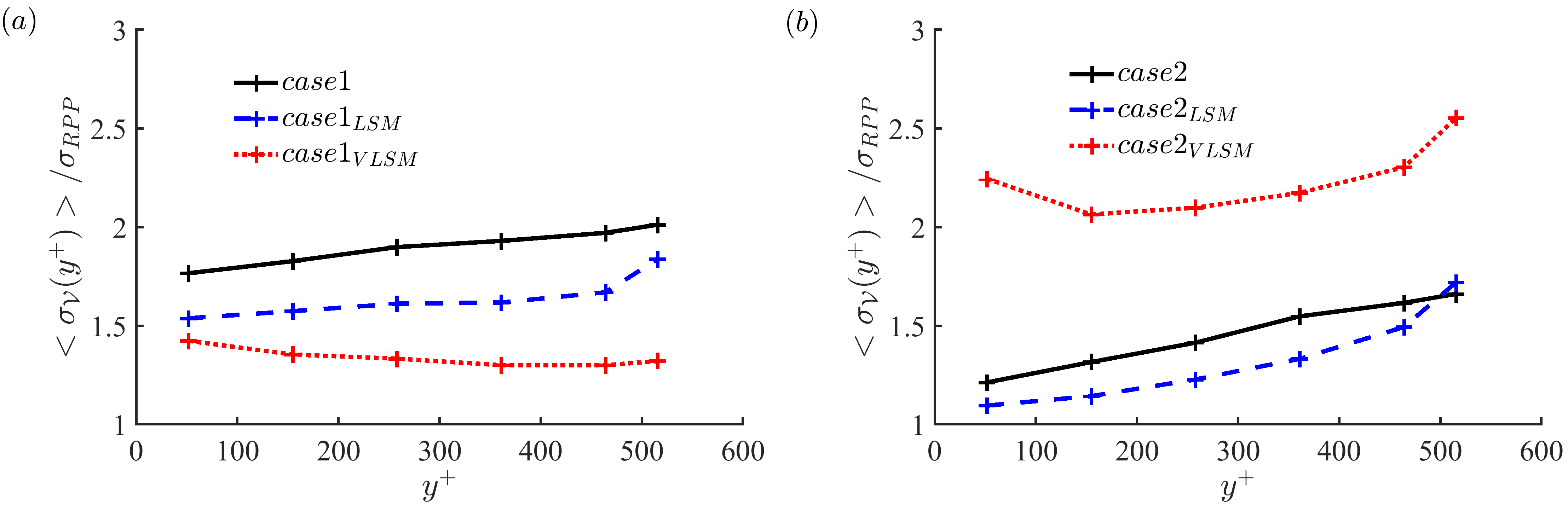}
\caption{\label{fig:voronoi_LSM_VLSM}Standard deviation of the normalized Vorono\"i area $\sigma_{\mathcal{V}}$, normalized by that of a random Poisson process, $\sigma_{RPP}$, as a function of height in wall-normal direction for (a) low and (b) high Stokes numbers coupled with different turbulent structures: full simulation ($case1,2$), LSMs ($case1,2_{LSM}$) and VLSMs ($case1,2_{VLSM}$).}
\end{figure*}

In order to quantify the particle clustering behavior shown in Fig. \ref{fig:Location_LSM_VLSM}, we again employ a Vorono\"i diagram analysis, shown in Fig. \ref{fig:voronoi_LSM_VLSM}. For low-inertia particles in Fig. \ref{fig:voronoi_LSM_VLSM}(a), the ratio $\sigma_\mathcal{V}/\sigma_{RPP}$ is highest in full simulation $case1$ but lowest in particles coupled only with VLSMs. In addition, $\sigma_\mathcal{V}/\sigma_{RPP}$ increases monotonically with increasing wall-normal distance in $case1$ and $case1_{LSM}$, while it decreases slightly in $case1_{VLSM}$ indicating a weak clustering when the particle/VLSM response time scale ratio ($St_{VLSM}=0.009$) is small. For high-inertia particles in Fig. \ref{fig:voronoi_LSM_VLSM}(b), the ratio $\sigma_\mathcal{V}/\sigma_{RPP}$ is slightly smaller in $case2_{LSM}$ than it is in full simulation $case2$. However, it is far larger in $case2_{VLSM}$ than it is in full simulation $case2$ due to the elongated structures which formed in $case2_{VLSM}$ and seen clearly in Fig. \ref{fig:Location_LSM_VLSM}(f).

%====================Figure================================================
\begin{figure*}
\includegraphics[width=16cm,trim={0cm 0cm 0cm 0cm}, clip]{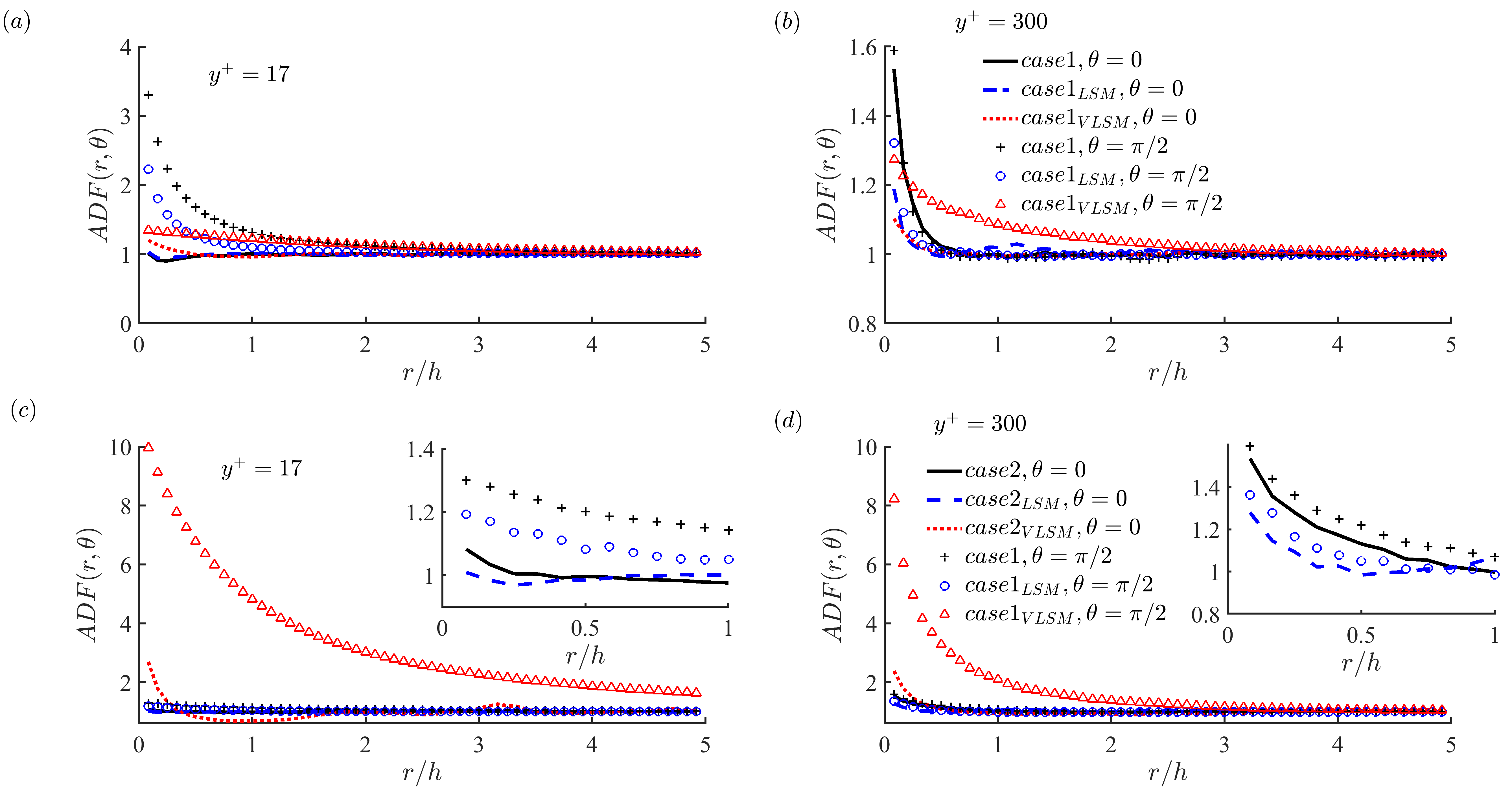}
\caption{\label{fig:RDF_LSM_VLSM} The streamwise and spanwise $ADF$ of particles in a slab with thickness of $2d_p$ at two wall-normal heights: near to the wall (a, c) $y^+=17$; in the outer layer (b, d) $y^+=300$, for two Stokes numbers: (a, b) low Stokes number; (c, d) high Stokes number. The insets of (c) and (d) show a zoom of small $ADF$ region within $r/h<1$. }
\end{figure*}

To gain insight into the anisotropic character of the particle clustering, the two-dimensional angular distribution functions are calculated as defined in Eq \ref{eq:ADF}, where particles are taken from a slab with thickness of $2d_p$: 

\begin{equation}\label{eq:ADF}
ADF(r,\theta)= \frac{\sum_{i=1}^{n_p} \delta N_i(r,\theta)/ (\delta r \cdot \delta \theta \cdot  n_p)}{N / (L_x \cdot L_y)},0 \leq \theta \leq \pi/2,
\end{equation} 

\noindent where $\delta N_i(r)$ is the particle number between $r - \delta r /2$ and $r + \delta r /2$ from the center of particle $i$, and $\delta N_i(r,\theta)$ is the particle number in a sector between $r - \delta r /2$ and $r + \delta r /2$ in the radial direction and $\theta - \delta \theta /2$ and $\theta + \delta \theta /2$ in the angular direction from the center of particle $i$; $\theta=0$ and $\theta=\pi/2$ correspond to the spanwise and streamwise directions, respectively. In the present study, we set $\delta r=0.08h$ ($\delta r^+=44$) and $\delta \theta=0.025\pi$ to compute $ADF(r,\theta)$. The mean value is from the average of $n_p$ particles from multiple snapshots in time. Finally, the distribution functions are normalized by the surface average particle number in $x-z$ plane ($n_p / L_x L_y$ representing a randomly distributed particle number density), where $n_p$ particles are from a two-dimensional $x-z$ slab taken in the wall-normal direction. Periodic boundary conditions are used for particles near the boundaries in the streamwise and spanwise directions. 

The $ADF(r,\theta)$ in the streamwise and spanwise directions corresponding to $\theta=\pi/2$ and $\theta=0$ at two different wall-normal heights ($y^+=17$ and $y^+=300$) are shown in Fig. \ref{fig:RDF_LSM_VLSM}. For low-inertia particles close to the wall, Fig. \ref{fig:RDF_LSM_VLSM}(a), the particle density from a reference particle in the streamwise direction is higher than in the spanwise direction, corresponding to the elongated anisotropic particle clustering formed in the inner layer as seen in Fig. \ref{fig:visu_LSM_VLSM}(a). Compared with the full simulation, the difference in the $ADF$ between the streamwise direction with the spanwise direction still exists in $case1_{LSM}$, whereas it diminishes in $case1_{VLSM}$. This indicates that the the elongated anisotropic particle clustering is similar between $case1_{LSM}$ and the full simulation (also Figs. \ref{fig:visu_LSM_VLSM}(c) and (a)) whereas the particle clustering tends to be more isotropic in $case1_{VLSM}$ (Fig. \ref{fig:visu_LSM_VLSM}(e)). For low Stokes number particles in the outer region, as shown in Fig. \ref{fig:RDF_LSM_VLSM}(b), the $ADF$ is similar between the streamwise direction and spanwise direction in both $case1$ and $case1_{LSM}$, which corresponds to the isotropic particle clustering formed in the outer layer as shown in Figs. \ref{fig:visu_LSM_VLSM}(a) and (b). In $case1_{VLSM}$, we see that the $ADF$ in the streamwise direction remains larger than unity even at a distance of $2.5h$ from the reference particle, which indicates that there are streamwise elongated structures formed (observed in Fig. \ref{fig:visu_LSM_VLSM}(b)), but not as pronounced as they are in the inner layer in $case1$. 

The $ADF$ of the high Stokes number particles is shown in Fig. \ref{fig:RDF_LSM_VLSM}(c, d). Compared with the full simulation, $case2_{VLSM}$ shows a significant increase of the streamwise $ADF$ and the sharp difference between the streamwise $ADF$ and the spanwise $ADF$. The presence in both Figs. \ref{fig:RDF_LSM_VLSM}(c, d) indicates that the elongated anisotropic particle clustering forms in both the inner layer and outer layer (see also Fig. \ref{fig:visu_LSM_VLSM}(f)). Meanwhile comparing $case2_{LSM}$ with the full simulation, the $ADF$ in the inner layer in both the streamwise and spanwise directions is slightly smaller than it is in the full simulation $case2$ as shown in Fig. \ref{fig:RDF_LSM_VLSM}(c), while the difference is small in the outer layer as shown in Fig. \ref{fig:RDF_LSM_VLSM}(d).

\subsubsection{Slip velocity} \label{subsubsec:feedback1_LSM_VLSM}

%====================Figure================================================
\begin{figure*}
\includegraphics[width=15cm,trim={0cm 0cm 0cm 0cm}, clip]{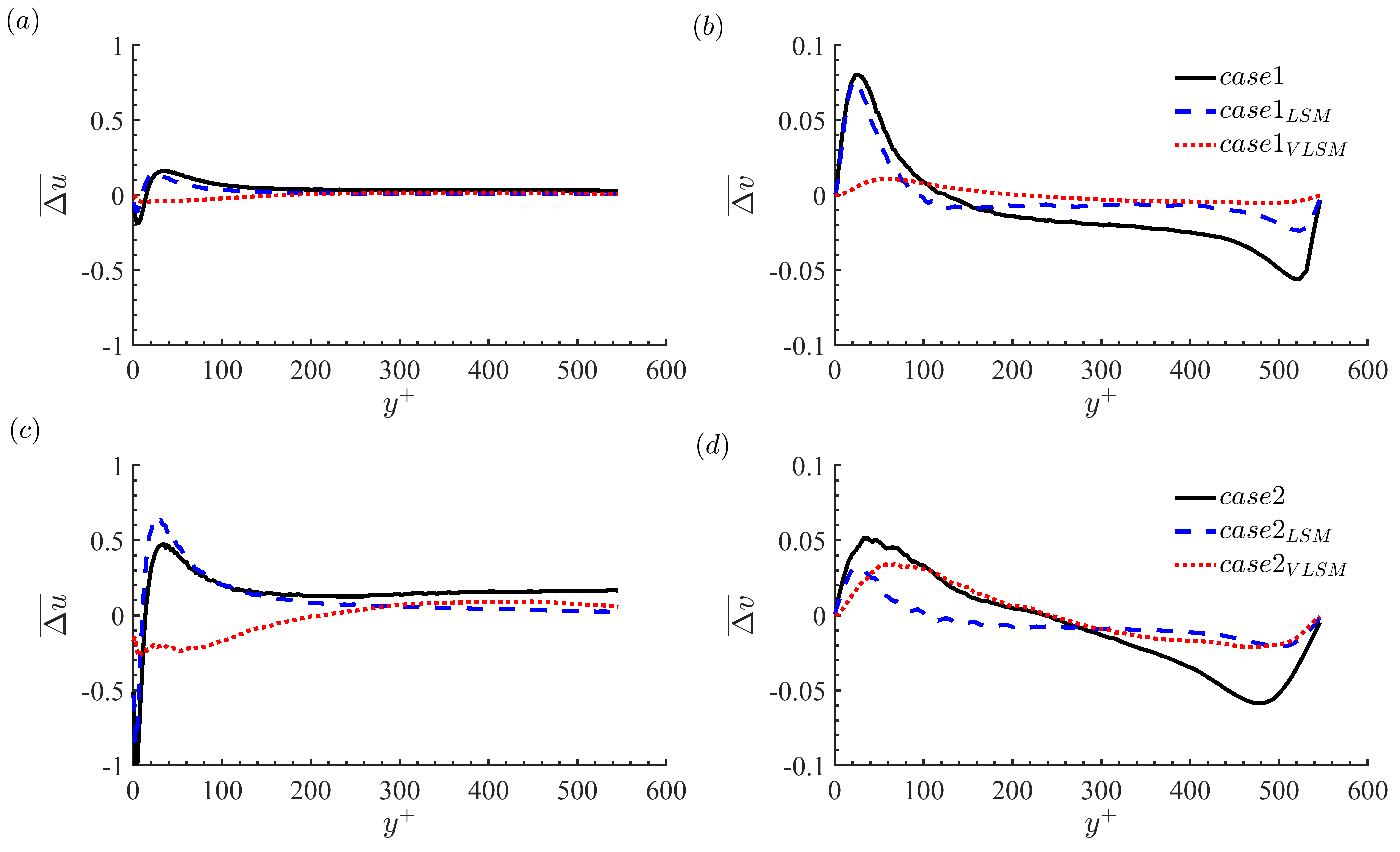}
\caption{\label{fig:Slip_velocity_LSM_VLSM} Profiles of the slip velocity $\Delta \mathbf{u}=\mathbf{u_f}-\mathbf{u_p}$. (a, c) streamwise velocity, $\overline{\Delta u}$; (b, d) wall-normal velocity, $\overline{\Delta v}$. (a, b) Low Stokes number; (c, d) High Stokes number.}
\end{figure*}

For particles with high inertia, a significant slip velocity ($\Delta \mathbf{u}=\mathbf{u_f}-\mathbf{u_p}$) can exist, which describes the exchange of momentum between the fluid and particle phases. A good prediction of the slip velocities is essential to predicting particle trajectories in particle-laden LES \cite{fede2006numerical}, Reynolds-averaged Navier-Stokes (RANS) coupled laden with Lagrangian particles \cite{arcen2009simulation}, or two-fluid modeling approaches \cite{simonin1993eulerian}. Through the slip velocity, the drag force governs the particle trajectories and segregation \cite{marchioli2002mechanisms}, and subsequently modulates the turbulent flow \cite{tanaka2008classification, zhao2013interphasial, wang_richter_2019}.

In the inner layer of turbulent channel flow, \citet{zhao2012stokes} find that in the streamwise direction the particles lead the fluid near the wall ($\overline{\Delta u}<0$ in $y^+<20$) whereas the particles lag behind the fluid away from the wall ($\overline{\Delta u}>0$ in $y^+>20$), and that the magnitude of the slip velocity increases monotonically with particle inertia. In the wall-normal direction, particles lag behind the fluid near the wall ($\overline{\Delta v}>0$ in $y^+<50$) whereas the particles lead the fluid away from the wall ($\overline{\Delta v}<0$ in $y^+>50$). As shown in Fig. \ref{fig:Slip_velocity_LSM_VLSM}, we find a similar trend as in \citet{zhao2012stokes} for both low and high inertia particles in the inner layer for the full simulations. In the outer layer ($y^+>100$), low-inertia particles tend to move towards the wall due to the negative mean drag force on the particle in the wall-normal direction. For high-inertia particles, the wall-normal slip velocity ($\overline{\Delta v}$) indicates that high-inertia particles drift towards the free surface (positive mean drag force exerted on the particle) not only in the inner layer, but also in the outer layer ($100<y^+<300$). Generally, in the inner layer, low-inertia particles coupled with LSMs ($case1_{LSM}$) produce the same sign and comparable magnitude of the slip velocity in the streamwise and wall-normal directions as shown in Fig. \ref{fig:Slip_velocity_LSM_VLSM}(a, b), respectively. However, high-inertia particles coupled with LSMs ($case2_{LSM}$) under-predict the $\overline{\Delta v}$ as shown in Fig. \ref{fig:Slip_velocity_LSM_VLSM}(d). In the outer layer, both of the artificial particle coupling tests with either low or high inertia particles tend to under-predict the magnitude of the wall-normal slip velocity. This suggests that the slip velocity is primarily due to particles coupling with LSMs, especially in the streamwise direction and for low inertial particles ($case1_{LSM}$). However, high-inertia particles coupling with VLSMs ($case2_{VLSM}$) are incorrect, even in the inner layer as shown in Fig. \ref{fig:Slip_velocity_LSM_VLSM}(d). This shows that in the inner layer the two-way coupling effect is mainly determined by LSMs for both low and high inertia particles especially in the streamwise direction (see Fig. \ref{fig:Slip_velocity_LSM_VLSM}(a, c)), whereas for high-inertia particles the wall-normal two-way coupling effect seems to be contributed by the transport of VLSMs (in Fig. \ref{fig:Slip_velocity_LSM_VLSM}(d)).

%As we have discussed before, the slip velocity is determined by the particle clustering in the corresponding turbulent structures and vice versa. 

\subsubsection{Particle feedback to the Reynolds stress budget} \label{subsubsec:feedback2_LSM_VLSM}

%====================Figure================================================
\begin{figure*}
\includegraphics[width=16.5cm,trim={0cm 0cm 0cm 0cm}, clip]{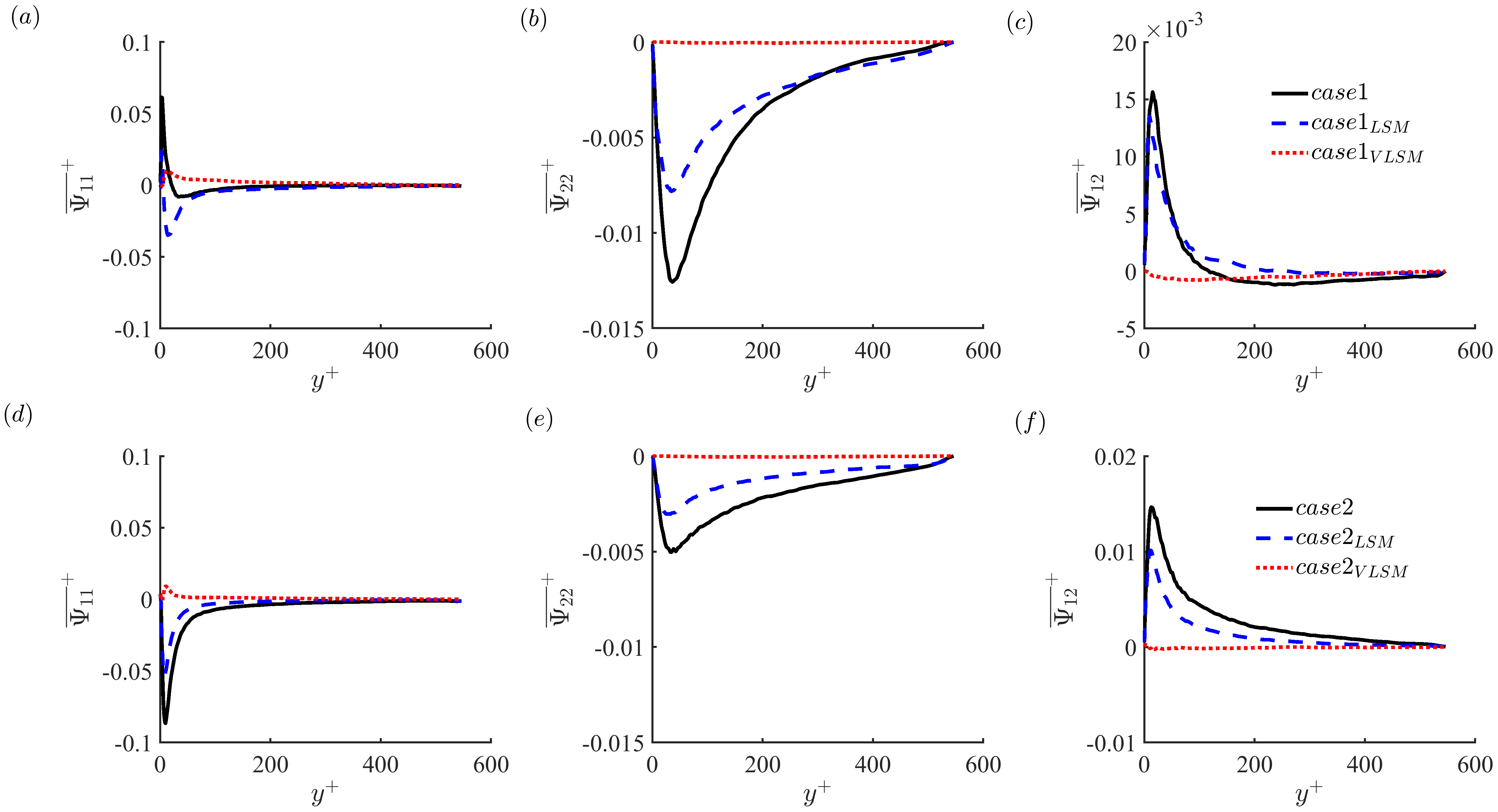}
\caption{\label{fig:Feedback_budget_LSM_VLSM} Profiles of the particle feedback terms to Reynolds stress budget, (a, d) particle sources to the $\overline{u'u'}$ budget, $\overline{\Psi_{11}}$; (b, e) particle sources to the $\overline{v'v'}$ budget, $\overline{\Psi_{22}}$; (c, f) particle sources to the $\overline{u'v'}$ budget, $\overline{\Psi_{12}}$. (a, b, c) Low Stokes number; (d, e, f) High Stokes number. All terms are scaled by $u_\uptau ^3/\delta$.}
\end{figure*}

The momentum exchange between the particle and fluid phases acts as a direct source/sink in the Reynolds stress budgets. Particle sources to the $\overline{u'u'}$, $\overline{v'v'}$ and $\overline{u'v'}$ budget are denoted as $\overline{\Psi_{11}}=\overline{F'_x u'}$, $\overline{\Psi_{22}}=\overline{F'_y v'}$ and $\overline{\Psi_{12}}=\overline{F'_x v' + F'_y u'}$, respectively \cite{wang2017modulation}. The particle sources are dependent on the characteristics of particle clusters \cite{capecelatro2018transition} and also strongly related to the particle inertia \cite{richter2015turbulence}. Furthermore, \citet{wang2019modulation, wang_richter_2019} demonstrated that both indirect and direct particle modulation mechanisms of LSMs and VLSMs have non-monotonic relationships with particle inertia, which can be observed by the particles' modulation of the Reynolds stress budgets in spectral space. Here we repeat a component of our previous analysis and show the three particle source terms $\overline{\Psi_{11}}$ (to the $\overline{u'u'}$ budget), $\overline{\Psi_{22}}$ ($\overline{v'v'}$ budget), and $\overline{\Psi_{12}}$ ($\overline{u'v'}$ budget) in Fig. \ref{fig:Feedback_budget_LSM_VLSM}. In general, inertial particles coupled with LSMs ($case1_{LSM}$ and $case2_{LSM}$) produce the same sign and comparable magnitude of the full particle sources, whereas the particle sources are nearly zero for the case of inertial particles coupled with VLSMs ($case1_{VLSM}$ and $case2_{VLSM}$). This shows that the particle sources to the Reynolds stress budgets are mainly dictated by the drag force interacting with small-scale structures (LSMs), which somewhat contradicts the direct enhancement mechanism of VLSMs by high inertial particles (the enhancement of VLSMs energy related to $\overline{\Psi_{12}}$ at high wavelengths in the outer layer) observed by \citet{wang_richter_2019}. In the other words, the direct enhancement mechanism of VLSMs cannot be captured simply by artificial coupling between high inertial particles and long-wavelength VLSM structures due to the underlying incorrect particle clustering in this case (seen in Figs. \ref{fig:visu_LSM_VLSM}(b) and (f)). It is clear that both turbulent structures (LSMs and VLSMs) work in tandem to simultaneously determine the correct particle clustering, which further works on the local fluid to modulate the turbulence. 

\subsubsection{Interphasial energy transfer and particle dissipation} \label{subsubsec:feedback3_LSM_VLSM}

Due to the slip velocity induced by particle inertia, we have shown in Sec. \ref{subsubsec:feedback2_LSM_VLSM} that the particles working on the fluid acts as the direct source/sink in the Reynolds stress budgets. At the same time, the drag force working on the particles represents the energy transferred from the fluid to the particles. The imbalance between the work transferred from the fluid to the particles with the particles to the fluid reflects energy dissipation which may help describe the mechanism of drag reduction in particle-laden flow \cite{zhao2013interphasial}. 

According to \citet{zhao2013interphasial}, the time rate of the work done by the local fluid to a particle $\dot{W_p}$, the work done by a particle on the local fluid $\dot{W_f}$, and the dissipation to heat $\epsilon$ is expressed as

\begin{align}\label{eq:energy_transfer}
\dot{W_p}&=6 \pi \mu a (u_{f,i}-u_{p,i}) u_{p,i}\\
\dot{W_f}&=-6 \pi \mu a (u_{f,i}-u_{p,i}) u_{f,i}\\
\epsilon &=\dot{W_p}+\dot{W_f}=-6 \pi \mu a (u_{f,i}-u_{p,i}) (u_{f,i}-u_{p,i})
\end{align} 

\noindent where $u_{p,i}$ and $u_{f,i}$ are the particle velocity and the fluid velocity seen by the particle, respectively.

%====================Figure================================================
\begin{figure*}
\includegraphics[width=16cm,trim={0cm 0cm 0cm 0cm}, clip]{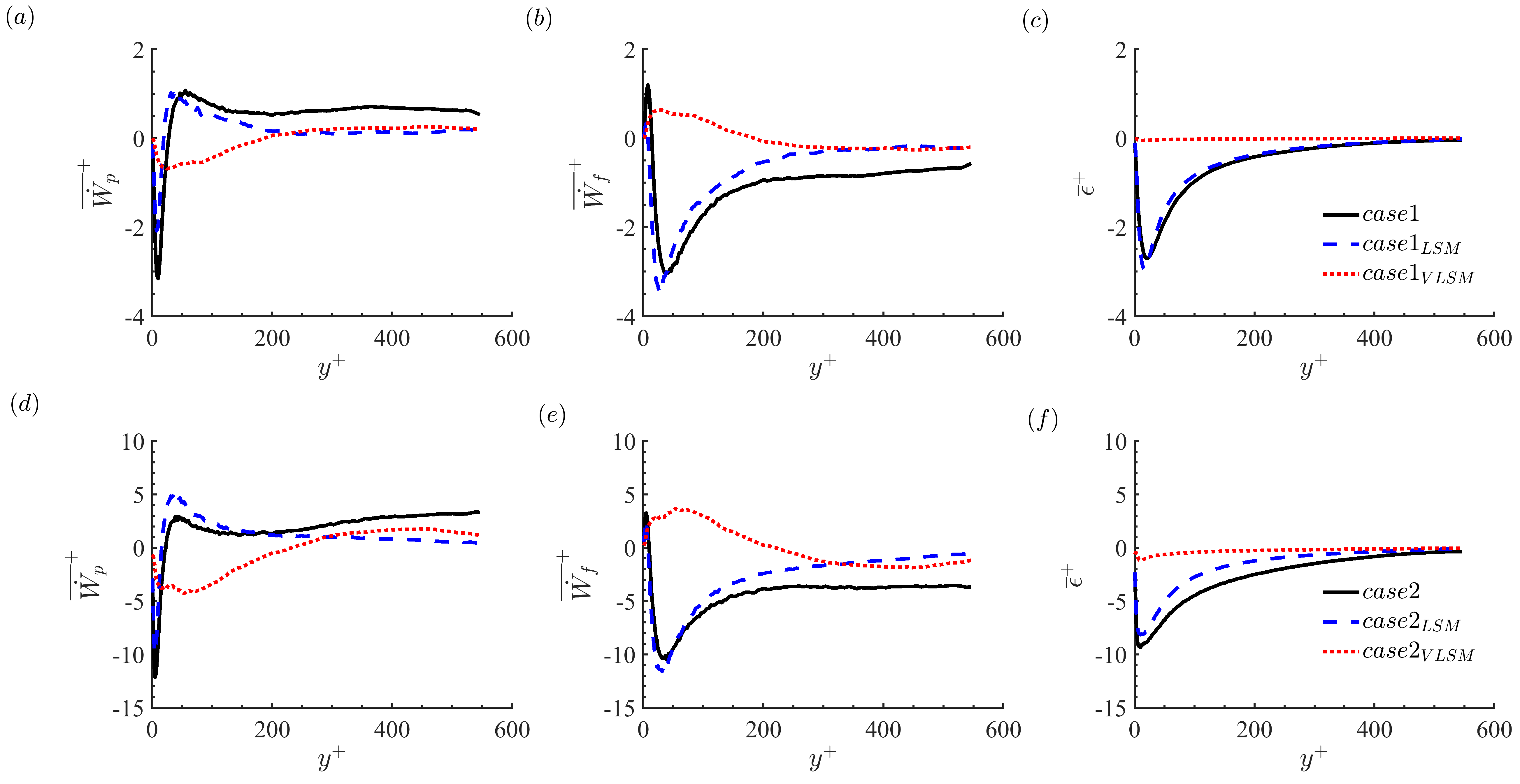}
\caption{\label{fig:Energy_transfer_LSM_VLSM} Profiles of the mean power transferred between fluid and particles: (a, d) from the fluid to the particle, $\overline{\dot{W_p}}$; (b, e) from the particle to the fluid, $\overline{\dot{W_f}}$; (c, f) the particle dissipation, $\overline{\epsilon}$. (a, b, c) Low Stokes number; (d, e, f) High Stokes number. The mean power is scaled by $6 \pi \mu a u_\uptau ^2$.}
\end{figure*}

The quantities $\dot{W_p}$, $\dot{W_f}$ and $\epsilon$ are shown in Figs. \ref{fig:Energy_transfer_LSM_VLSM}(a, d), (b, e) and (c, f) for low and high Stokes number particles, respectively. In the inner layer, the sign and trend of $\dot{W_p}$, $\dot{W_f}$, and $\epsilon$ profiles in full simulations ($case1$ and $case2$) are qualitatively similar to those obtained by \citet{zhao2013interphasial} at $Re_\uptau=180$. The particles exert work on the local fluid in the buffer layer and viscous layer ($\overline{\dot{W_p}}<0$, $\overline{\dot{W_f}}>0$), whereas the particles receive energy from the fluid ($\overline{\dot{W_p}}>0$, $\overline{\dot{W_f}}<0$) beyond $y^+=40$. The energy transfer between the particles and the fluid is nearly the same between the particles coupled with LSMs ($case1_{LSM}$ and $case2_{LSM}$) and the full simulations ($case1$ and $case2$). However, large differences are seen between the VLSM coupling cases and the full simulations, not only in the magnitude, but also in sign. In the outer layer, both the low and high inertia particles continuously receive energy from the large-scale fluid motions ($\overline{\dot{W_p}}>0$, $\overline{\dot{W_f}}<0$), but the magnitude is smaller in both artificial coupling tests than it is in full simulation. As shown in Fig. \ref{fig:Energy_transfer_LSM_VLSM}(c, f), across the whole wall-normal height, the particle dissipation is comparable between particles coupled with LSMs and the full simulation whereas $\overline{\epsilon}$ is negligible in both $case1_{VLSM}$ and $case2_{VLSM}$. This confirms that the particle dissipation generally comes from particles coupling with low-wavelength structures.

From the above discussions regarding the slip velocity (Fig. \ref{fig:Slip_velocity_LSM_VLSM}), particle feedback to the Reynolds stress budget (Fig. \ref{fig:Feedback_budget_LSM_VLSM}), and interphasial energy transfer (Fig. \ref{fig:Energy_transfer_LSM_VLSM}), a similar conclusion can be drawn that the two-way coupling effect is mainly due to particles interacting with LSMs, especially in the inner layer, for both low and high inertia particles. The two-way coupling effect is rather small in case of only coupling with VLSMs in both the inner layer and outer layer, suggesting that while VLSMs are important for distributing particles throughout the domain, and although their strength can be modulated by particles \citep{richter2015turbulence,wang_richter_2019}, it is fundamentally the coupling between LSMs and particles which dictate energy and momentum transfer between phases, even for high-inertia particles. 

%==============================================================================
\section{Conclusion} \label{sec:Conclusion}

In this study, we investigate the transport of inertial particles by large-scale motions (LSMs) and very-large-scale motions (VLSMs) in moderate Reynolds number ($Re_{\uptau}=550$) open channel flow. Two particle Stokes numbers based on the characteristic time scales of the LSMs and VLSMs are used, where low-inertia particles with $St_{LSM}=0.0625$ preferably accumulate in LSMs in the inner layer \cite{wang2019modulation} and high-inertia particles with $St_{VLSM}=0.069$ tend to form particle clustering structures associated with VLSMs in the outer layer \cite{wang_richter_2019}.

The first test uses a truncated domain size to isolate VLSMs because the VLSMs can only be captured in a sufficiently large domain. By comparing the flow field between a small domain and large domain in single-phase flow, it is confirmed that the small domain can capture the correct length and intensity of LSMs in spectral space within the inner layer, even though the VLSMs contribute $7-20\%$ of the Reynolds shear stress and $10-35\%$ of the turbulent kinetic energy in the large domain simulation. As a consequence, an increase of wall particle concentration is found in the large domain simulation, up to $20\%$ different with respect to the small domain simulation, which is similar as previously investigated by \citet{sardina2012wall} at low $Re_{\uptau}=180$. From a Vorono\"i tessellation analysis, the particles' preferential concentration is higher in the large domain simulation than it is in small domain simulation. However, the LSMs still correlate with each other in small domain simulation, particularly in the streamwise direction. This effect cannot be excluded in this test \cite{sardina2012wall}, even though the VLSMs effect is considered to be more important than the effect of decorrelated LSMs on the particle distribution. 

In order to exclude the effect of correlated LSMs, we then perform an artificial coupling test between a filtered flow field (i.e. to isolate LSMs and VLSMs) with inertial particles in a large domain size, in order to compare with the full simulation. Similar to the truncated domain size effect, the particle concentration and the underlying turbophoresis are under-predicted when VLSMs are absent. The particle preferential concentration is more closely related to LSMs than VLSMs for both kinds of particles ($St_{LSM}=0.0625,0.475$) as seen from Vorono\"i tessellation analysis. From a two-dimensional angular distribution function analysis, for low-inertia particles coupling with VLSMs ($St_{VLSM}=0.009$), particle clustering is more isotropic than in full simulation in the inner layer whereas weak, elongated streamwise anisotropic structures are formed in the outer layer. For high-inertia particles coupled with VLSMs ($St_{VLSM}=0.069$), strong, elongated streamwise anisotropic structures are formed in both the inner layer (typical streamwise scale is longer than $5h$) and the outer layer (typical streamwise scale is around $2h$). These large-scale organized particle structures induced by VLSMs are shorter and less organized than observations in turbulent plane Couette flow at low $Re_\uptau=167$ \cite{bernardini2013effect}. 

These findings have implications on the ability of developing subgrid models for particle two-way coupling in LES. The transfer of energy from particles to/from the fluid, on one hand, is mainly due to particles interacting directly with LSMs, especially for particles in the inner layer, whereas the two-way coupling effect is rather small when coupled directly with VLSMs in both the inner layer and outer layer. However, this unfortunately does not mean that the effects of VLSMs can be ignored, since their energy content and contribution to the Reynolds stress can be altered by this two-way coupling, and it was also observed in this study that VLSMs alone are an integral part of the spatial distribution of particles. Properly representing these effects remains an ongoing challenge in multiphase LES.

%A posteriori knowledge of the subgrid model effect on the particle two-way coupled LES in the inner layer and the sensitivity of particle Stokes number characterized by the Kolmogorov scale can refer to \citet{vreman2009two} and \citet{fede2006numerical}, respectively. Furthermore, concerning about the two-way coupling effect with VLSMs, the slip velocity, particle feedback to the Reynolds stress budget and interphasial energy transfer and particle dissipation are analyzed. The similar conclusion can be drawn that the two-way coupling effect is mainly due to particles interact with LSMs especially for particles in the inner layer whereas the two-way coupling effect is rather small in case of only coupling with VLSMs in both the inner layer and outer layer. This suggests that the importance of capturing LSMs during two-way particle-laden LES simulation, even though \citet{hwang2010self} find that in single-phase flow simulation the VLSMs can be reproduced by LES by quenching the LSMs. 

%==============================================================================
\begin{acknowledgments}
The authors acknowledge grants G00003613-ArmyW911NF-17-0366 from the US Army Research Office and N00014-16-1-2472 from the Office of Naval Research. Computational resources were provided by the High Performance Computing Modernization Program (HPCMP), and by the Center for Research Computing (CRC) at the University of Notre Dame.
\end{acknowledgments}

%==============================================================================
%\appendix
%
%\section{Appendixes A}

% The \nocite command causes all entries in a bibliography to be printed out
% whether or not they are actually referenced in the text. This is appropriate
% for the sample file to show the different styles of references, but authors
% most likely will not want to use it.
%\nocite{*}

\bibliographystyle{apsrmp4-2}
\bibliography{Gwang_PRF}
%\bibliography{/home/gwang4/Documents/Postdoc_ND/gwang_JabRef/SINGLE-PHASE,/home/gwang4/Documents/Postdoc_ND/gwang_JabRef/postdoc_ND,/home/gwang4/Documents/Postdoc_ND/gwang_JabRef/TWO-PHASE_JAB}% Produces the bibliography via BibTeX.
\end{document}